\pdfoutput=1
\documentclass[aps,pre,amssymb,amsmath]{revtex4}
\usepackage{floatflt}
\usepackage{graphicx}
\usepackage{hhline}

\newcommand{\be}{\begin{equation}}
\newcommand{\ee}{\end{equation}}
\newcommand{\bea}{\begin{eqnarray}}
\newcommand{\eea}{\end{eqnarray}}
\newcommand{\p}{\partial}
\newcommand{\vvect}{\mathbf{v}}
\newcommand{\mydiv}{\mathrm{div}\,}
\newcommand{\mynabla}{\pmb{\nabla}}
\newcommand{\curl}{\pmb{\nabla\times}\,}

\begin{document}
\title{Dependence of fluid flows in an evaporating sessile droplet 
on the characteristics of the substrate}
\author{L.Yu.\ Barash\,$^{1,2,3,a)}$}
\affiliation{
$^{1)}$ Landau Institute for Theoretical Physics, 142432 Chernogolovka, Russia \\
$^{2)}$ Moscow Institute of Physics and Technology, 141700 Moscow, Russia \\
$^{3)}$ Science Center in Chernogolovka, Russian Academy of Sciences, 142432 Chernogolovka, Russia \\
e-mail: \tt $^{a)}$barash@itp.ac.ru}
\begin{abstract}
Temperature distributions and the corresponding vortex structures 
in an evaporating sessile droplet are obtained by performing detailed 
numerical calculations. A Marangoni convection 
induced by thermal conduction in the drop and the substrate
is demonstrated to be able to result not only in a single vortex, but also in 
two or three vortices, depending on the ratio of substrate to fluid 
thermal conductivities, on the substrate thickness and the contact 
angle. The ``phase diagrams'' containing information on the number, 
orientation and spatial location of the vortices 
for quasistationary fluid flows are presented and analysed.
The results obtained demonstrate 
that the fluid flow structure in evaporating droplets can be 
influenced in a controlled manner by selecting substrates with
appropriate properties.
\end{abstract}
\maketitle

\section{Introduction}

It is known that the evaporating flux density along
the surface of a drying sessile droplet is inhomogeneous 
and diverges on approach to the pinned contact line~\cite{Deegan97,Deegan2000}.  
The evaporation can induce temperature variations in
the vicinity of the substrate--liquid interface and along the drop surface.
A temperature variation along the liquid--vapor interface can in turn
generate a thermocapillary flow inside the drop which
has been intensively studied~(see, for example,
the review articles~\cite{Erbil,Larson} and references therein).
Sessile drop evaporation processes and structures of the fluid flow
are of interest for important applications 
in ink-jet printing~\cite{Park2006,Calvert2001}, spraying of
pesticides~\cite{Zhu2009}, micro/nano fabrication~\cite{Xia2008,Caroll2006}, 
thin film coatings~\cite{Kimura2003}, biochemical assays~\cite{Nguyen2002}, 
spray cooling~\cite{Jia2003}, disease diagnosis~\cite{Brutin2011,Trantum2012}, 
deposition of DNA/RNA micro-arrays~\cite{Schena1995,Fang2006,McHale2007}, 
and manufacture of novel optical and electronic materials~\cite{Kawase2001}.

The substrate temperature
distribution under the drop can be measured using
thermo-chromic liquid crystals~\cite{Sodtke2007} and with IR thermography~\cite{Tarozzi2007}. 
The substrate properties play an important role
in the nanotechnology applications of the problem.
One of the examples is self-assembly of superlattices of nanoparticles
taking place during evaporation of colloidal solutions.
It is known that substrate characteristics can strongly influence 
both the deposition patterns and the self-assembly 
process~\cite{Voylov,Wang,Ge,Cai,Lin2,Bigioni}.

Numerical calculations of Marangoni convection
in an axially symmetrical evaporating droplet
agree well with corresponding experimental data~\cite{HuLarsonReverse,SavinoFico}.
A sensitivity of Marangoni fluid flows to the droplet contact angle 
is known since Hu and Larson demonstrated that fluid circulation
in the vortex can reverse its sign at a critical contact angle 
for a drop placed onto substrates with finite thicknesses~\cite{HuLarsonMarangoni}.
It was originally observed and described by Ristenpart et al. 
that the circulation direction depends on the substrate to liquid 
ratio of the thermal conductivities~\cite{Ristenpart}.
Specifically, the authors found that the ratio determines the sign 
of the tangential component of the temperature gradient 
at the surface close to the contact line, and, therefore, 
it determines a direction of the circulation in that region.
Assuming the key role of a small vicinity of the contact line in forming the 
circulation direction in a single vortex, the shape of the liquid--vapor 
interface in that small region was approximated as a plane,
which is actually inappropriate since the three-phase
contact line has finite radius of curvature.
However, the conditions for the circulation sign change 
have been found within such framework.
While the approach and predictions of Ref.~\cite{Ristenpart} 
generally are qualitatively insightful and 
quite useful, the particular approximations made and the 
corresponding quantitative results obtained in~\cite{Ristenpart} have not been 
justified by more accurate numerical calculations.
For example, it follows from the analytical results of~\cite{Ristenpart}
that for $k_R<1.45$ and $k_R>2$ the circulation direction is insensitive
to the contact angle. The results of more recent numerical calculations,
as well as the results of the present study, do not confirm this statement.

An alternative approach suggested by Xu et al.~\cite{Xu}, focuses on
a heat transfer in the 
immediate vicinity of the symmetry axis piercing the apex. The
change of sign of the tangential gradient of the temperature near the
apex and, hence, the corresponding transition between the opposite 
circulation directions, taking place with a variation of the relative 
substrate--liquid thermal conductivity, has been identified by the 
authors under different conditions as compared to the results of~\cite{Ristenpart}.
Transition points at various contact angles obtained
in~\cite{Xu} are much closer to the results of subsequent 
numerical calculations and are in agreement with the corresponding experimental data.

The above studies~\cite{HuLarsonMarangoni,Ristenpart,Xu} assumed a monotonic temperature profile 
along the droplet surface, and, hence, a single-vortex fluid flow. 
However, such an assumption only partially explains the phenomenon.
The thermal conduction processes throughout the droplet
can generally result in a nonmonotonic spatial dependence of the surface 
temperature and in more complicated convection patterns inside a drop.
In particular, either a single vortex or several vortices are formed 
in the droplet depending on the thermal conductivity of the substrate~\cite{CCP2011}.
With varying the relative substrate--liquid thermal conductivity,
transitions between regimes with different numbers of vortices and/or
circulation directions take place.
This has been described recently in more detail by Zhang et al. in~\cite{Zhang},
where the authors presented the ``phase diagram'' characterizing, for
a fixed substrate thickness $h_S=0.1 R$,
the distribution of surface temperature in the $k_R$--$\theta$ plane,
where the notations $k_R$, $R$ and $\theta$ are explained in Table~\ref{ParamTable}.

The three regions in the $k_R$--$\theta$ plane have been demonstrated 
in~\cite{Zhang}. In region I the surface temperature monotonically increases 
from the center to the edge of the droplet. In region III the surface 
temperature decreases monotonically from the center to the edge of the 
droplet. Finally, in region II the temperature exhibits a nonmonotonic spatial 
dependence along the droplet surface. Naturally, regions I and III 
correspond to single-vortex states with opposite circulations of the 
fluid flows. In the present work we focus on the substructure of the region II.
Specifying the number of the temperature extrema along the 
droplet surface, we find the subregions, which correspond to two or 
three vortices inside the droplet. We also identify the dependence of
the borders between the subregions on the substrate thickness.
The results obtained demonstrate that the vortex state structure in
evaporating droplets of capillary size can be prepared in a controlled manner
by selecting substrates with appropriate thermal conductivity 
and thickness.

\begin{table}
\caption{ The notations and the parameter values used for
obtaining the evaporation rates, temperature distribution 
and hydrodynamics in the drop.
The tabular data are taken from~\cite{Lide}.}
\begin{tabular}{|c|l|l|}
\hline
Drop & Initial temperature        & $T_0=293.15$ K \\
parameters  & Contact line radius             & $R=10^{-3}$ m \\
& Contact angle & $\theta$ \\
\hline
\hline
Substrate & Radius & $R_S=1.25\cdot 10^{-3}$ m \\
parameters & Thickness & $h_S$ \\
& Ratio of substrate thickness to contact line radius & $h_R=h_S/R$ \\
& Thermal conductivity & $k_S$ \\
& Substrate to liquid ratio of thermal conductivities & $k_R=k_S/k_L$ \\
\hline
\hline
Fluid
&Density                & $\rho=813.6$ kg/m$^3$ \\
characteristics
& Molar mass & $\mu=0.10217$ kg/mole\\
(1-hexanol)&Thermal conductivity   & $k_L=0.15 $ W/(m$\cdot$ K) \\
&Thermal diffusivity    & $\kappa=k_L/(\rho c_p)=7.84\cdot 10^{-8}$ m$^2$/s \\
&Dynamic viscosity      & $\eta=4.578\cdot 10^{-3}$ kg/(m$\cdot$ s) \\
&Surface tension        & $\sigma=0.02581$ kg/s$^2$ \\
&Temperature derivative of surface tension  & $-\p\sigma/\p T=8.0\cdot 10^{-5}$\ kg/(s$^2\cdot$ K) \\
&Latent heat of evaporation  & $L=6.03\cdot 10^5$ J/kg \\
\hline
\hline
1-hexanol vapor
&Diffusion constant & $D=6.21\cdot 10^{-6}$ m$^2$/s\\
characteristics
&Saturated 1-hexanol vapor density & $u_s=6.55\cdot 10^{-3}$ kg/m$^3$ \\
\hline
\hline
\end{tabular}
\label{ParamTable}
\end{table}

\section{Basic equations and methods}
\label{MethodsSec}

The basic hydrodynamic equations inside the drop are the Navier--Stokes equations
and the continuity equation for the incompressible fluid
\begin{eqnarray}
\frac{\p\vvect}{\p t}+(\vvect\cdot\mynabla)\vvect+
\frac1\rho\,{\mathrm{grad}\,p}
&=&\nu\,\Delta{\vvect},
\label{NavSt}
\\
\mydiv\vvect&=&0.
\label{ZeroDiv}
\end{eqnarray}
Here $\Delta=\p^2/\p r^2+\p/r\p r+\p^2/\p z^2$,
$\nu=\eta/\rho$ is kinematic viscosity,

The calculation of thermal conduction inside
the droplet and the substrate
is carried out without taking into account
the convective heat transfer, which
is justified when the P\'eclet number 
$Pe=\overline{u}R/\kappa$ is much smaller
than unity. The following equation is solved:
\begin{equation}
\frac{\p T}{\p t} = \kappa \Delta T,
\label{EqThermal}
\end{equation}
where $\kappa=k/(\rho c_p)$ is thermal diffusivity.
The boundary conditions take the form $\p T/\p r=0$ for $r=0$;
${\p T}/{\p n}=-{Q_0(r)}/{k}$ at the drop surface.
Here
$Q_0(r)=LJ_s(r)$ is the rate of heat loss per unit area
of the upper free surface,
$\mathbf{n}$ is a normal vector to the drop surface,
$J_s$ is the local evaporation rate determined by
\be
J_s(r)=J_0(\theta)(1-r^2/R^2)^{-\lambda(\theta)},
\label{DeeganFit}
\ee
where $\lambda(\theta)=1/2-\theta/\pi$,
fitting expression for $J_0(\theta)$ is taken from~\cite{HuLarsonEvap},
other notations are explained in Table~{\ref{ParamTable}}.

The relation~(\ref{DeeganFit}) fits well with the analytical solution
for a stationary spatial distribution of the vapor concentration
for a drop with the shape of a spherical cap~(see~\cite{Deegan2000,HuLarsonEvap}).
The problem is mathematically equivalent to that solved by
Lebedev~\cite{Lebedev}, who obtained the electrostatic potential
of a charged conductor having the shape defined by two
intersecting spheres. 
Nonstationary effects in vapor concentration
and effects resulting from deviations of droplet shape from spherical cap 
are considered in detail in~\cite{Barash2009}. 
Nonstationary effects in vapor concentration
may only effectively result in a change of 
the constant $J_0(\theta)$ in~(\ref{DeeganFit})
and become small for $t\gg R^2/D$, where $t$ is the duration
of the evaporation process.
Effects resulting from deviations of a droplet shape from spherical cap 
are very small provided that the Bond number $B_o=\rho g h R/(2\sigma\sin\theta)$ 
is much smaller than unity, which is true for the droplets considered.
Evaporation rate measurements for such droplets agree well with the calculations
based on the relation~(\ref{DeeganFit})~\cite{Deegan2000,HuLarsonEvap,Barash2009}.

The relation $Q_0(r)=LJ(r)$ implies that the heat flow
from ambient air towards the drop surface is negligible.
This is the case if the temperature difference between
drop surface and the air far from the drop
is less than $LDu_s/k$~\cite{Fuchs}.
This is well satisfied in the problem in question.
The substrate radius is $5R/4$ (see also Table~\ref{ParamTable}).
At the substrate--fluid interface we have the matching condition
$k_S \p T_S/\p z= k_L \p T_L/\p z$.
At the substrate--gas interface we have $\p T_S/\p n=0$,
where $\mathbf{n}$ is a normal vector to the substrate surface,
since the thermal conductivity of the air is negligibly small.
At the bottom of the substrate we have the boundary
condition $T_S=T_0$.

Taking the curl of both sides of equation (\ref{NavSt}),
one excludes the pressure $p$ and obtains
\begin{equation}
\frac{\p}{\p t}(\curl\vvect) + (\vvect\cdot\mynabla)(\curl\vvect)
-((\curl\vvect)\cdot\mynabla)\vvect =\nu \Delta(\curl \vvect).
\end{equation}
Therefore,
\begin{equation}
\frac{\p}{\p t}\gamma(r,z)+(\vvect\cdot\mynabla)\gamma(r,z)=
\nu \left(\Delta\gamma(r,z)-\frac{\gamma(r,z)}{r^2}\right),
\label{EqGamma}
\end{equation}
where the vorticity $\gamma$ is introduced by
\begin{equation}
\gamma(r,z)=\frac{\p v_r}{\p z}-\frac{\p v_z}{\p r};\qquad
\curl\vvect=\gamma(r,z)\mathbf{i}_\varphi.
\end{equation}
We define the stream function $\psi$, such that
\begin{equation}
\frac{\p\psi}{\p z}=rv_r\ , \qquad
\frac{\p\psi}{\p r}=-rv_z
\label{Psi}
\end{equation}
One has $\p^2\psi/(\p r\p z)=v_r+r(\p v_r/\p r)=-r(\p v_z/\p z)$,
therefore $v_r/r +{\p v_r}/{\p r}+{\p v_z}/{\p z}=0$, i.e.
the velocities obtained with Eqs.(\ref{Psi})
will automatically satisfy the requirement (\ref{ZeroDiv}).

Applying the Laplace operator to the stream function,
one obtains $\Delta\psi=r\gamma-2v_z$, which can be transformed
to a more convenient form $\tilde\Delta\psi=r\gamma$ with
the modified operator $\tilde\Delta$ that differs from
the Laplace operator by the sign of the term $\p/(r\p r)$:
\begin{equation}
\tilde\Delta\psi=
\frac{\p^2\psi}{\p r^2}-\frac1{r}\frac{\p\psi}{\p r}+
\frac{\p^2\psi}{\p z^2}=r\left(\frac{\p v_r}{\p z}-
\frac{\p v_z}{\p r}\right)=r\gamma.
\label{LaplacePsi}
\end{equation}

The method of the numerical solution is as follows:
(1) we solve Eq.~(\ref{EqThermal}) and obtain the
temperature distribution inside the droplet and the substrate,
which, in particular, allows us to obtain the boundary conditions 
for Eq.~(\ref{EqGamma});
(2) we solve Eq.~(\ref{EqGamma}) with the proper boundary conditions
and find $\gamma(r,z)$;
(3) we solve Eq.~(\ref{LaplacePsi}) and find $\psi(r,z)$.
Then we find velocities with Eqs.(\ref{Psi}) and specify
the boundary conditions for $\gamma(r,z)$.
We repeat the steps 2 and 3 until the velocities 
become constant.
The proper boundary conditions for quantities $\gamma$ and $\psi$,
satisfying Eqs.~(\ref{EqGamma}) and (\ref{LaplacePsi}),
take the form
$\gamma=0$ for $r=0$; $\gamma=\p v_r/\p z$ for $z=0$;
$\gamma={d\sigma}/(\eta ds)+2v_\tau d\phi/ds$
on the surface of the drop; $\psi=0$ at all boundaries: at the
surface of the drop, at the axis of symmetry
of the drop, and at the bottom of the drop
(see Appendix~B in~\cite{Barash2009}).
Here ${d\sigma}/{ds}=-\sigma'\p T/\p s$ is the derivative
of surface tension along the surface of the drop,
where the distribution of temperatures, which is found with
Eq.~(\ref{EqThermal}) is taken into account, and
$\sigma(T)=\sigma_0-\sigma'(T-T_0)$
is the experimental dependence of surface tension $\sigma$ on
temperature $T$.

In order to accurately take into account 
the vicinity of the three-phase contact line,
which is a very important area of calculation, 
we have used implicit finite 
difference method and irregular mesh inside the droplet for
the calculation of thermal conduction.
We use the following mesh in the drop and the substrate:
$r_i=R\left(1-(n-i)^2/n^2\right)$, $i=0,\dots,n$; 
$r_i=R\left(1+(i-n)^2/n^2\right)$, $i=n+1,\dots,3n/2$; 
$z_j=hj^2/n^2$, $j=0,\dots,n$,
where $h$ is the droplet height and $n=200$.
Inside the substrate, we use the following mesh:
$r_{Si}\equiv r_i$, $i=0,\dots,n$; 
$z_{Sj}=jh_S/n$, $j=0,\dots,n$, where $h_S$ is the
substrate thickness.
Such irregular mesh allows to substantially
increase the accuracy of the calculation in the
vicinity of the contact line.

We denote the distances between the mesh point $(i,j)$
and its nearest neighbors as
$a=r_i-r_{i-1}$, $b=r_{i+1}-r_i$, $c=z_{j+1}-z_j$ and
$d=z_j-z_{j-1}$. Then the finite difference representations
for the second derivatives are

\begin{eqnarray}
\hat\delta_r^2 T=\frac2{(a+b)ab}\left(aT_{i+1,j}^{n}-(a+b)T_{ij}^{n}+bT_{i-1,j}^{n}\right),\\
\hat\delta_z^2 T=\frac2{(c+d)cd}\left(dT_{i,j+1}^{n}-(c+d)T_{ij}^{n}+cT_{i,j-1}^{n}\right),
\end{eqnarray}
where $T_{ij}^{n}$ is the temperature at mesh point $(i,j)$ for the $n$th time step.

We apply the alternating direction method
to Eqn.~(\ref{EqThermal}) with the above notations.
In the first part of the method one takes
$r$-derivative implicitly. Then the finite difference
representation of Eqn.~(\ref{EqThermal}) is
\be
\frac{T_{ij}^{n+1/2}-T_{ij}^n}{\kappa h_t/2}=
\hat\delta_r^2 T_{ij}^{n+1/2}+\hat\delta_z^2 T_{ij}^{n}+
\frac{1}{r}\frac{T_{i+1,j}^{n+1/2}-T_{i-1,j}^{n+1/2}}{a+b}.
\ee
For given temperature at time step $n$ it is
convenient to rewrite this expression as
\be
c_i'T_{i-1,j}^{n+1/2}+(d_i'-2/(\kappa h_t))T_{ij}^{n+1/2}
+e_i'T_{i+1,j}^{n+1/2}=
c_j''T_{i,j-1}^n+(d_j''-2/(\kappa h_t))T_{ij}^n+e_j''T_{i,j+1}^n.
\label{implicit1}
\ee
Here
\bea
c_i'=2/((a+b)a)-1/((a+b)r),& d_i'=-2/(ab),&
e_i'=2/((a+b)b)+1/((a+b)r),
\label{koeff1}
\\
c_j''=-2/((c+d)d),& d_j''=2/(cd),& e_j''=-2/((c+d)c),
\\
c_0'=0,& d_0'=-4/a^2,& e_0'=4/a^2,
\label{koeff4}
\eea
For each $j\ne 0$ the tridiagonal matrix algorithm is used
to solve the set of equations (\ref{implicit1})
for the temperature at the time step $n+1/2$.
The boundary interpolation at the droplet surface (see below) and
the boundary condition at the substrate--gas interface
are solved together with the set of equations (\ref{implicit1})
by the tridiagonal matrix algorithm.

\begin{figure}[tb]
\caption{Temperature distribution, distribution of absolute value of velocity 
and vector field plot of velocity for a droplet of 1-hexanol in a single-vortex
regime obtained for $\theta=35^\circ$, $k_R=1$, $h_R=h_S/R=0.05$.}
\hspace*{-0.1cm}
\raisebox{0.8cm}{\includegraphics[width=0.35\textwidth]{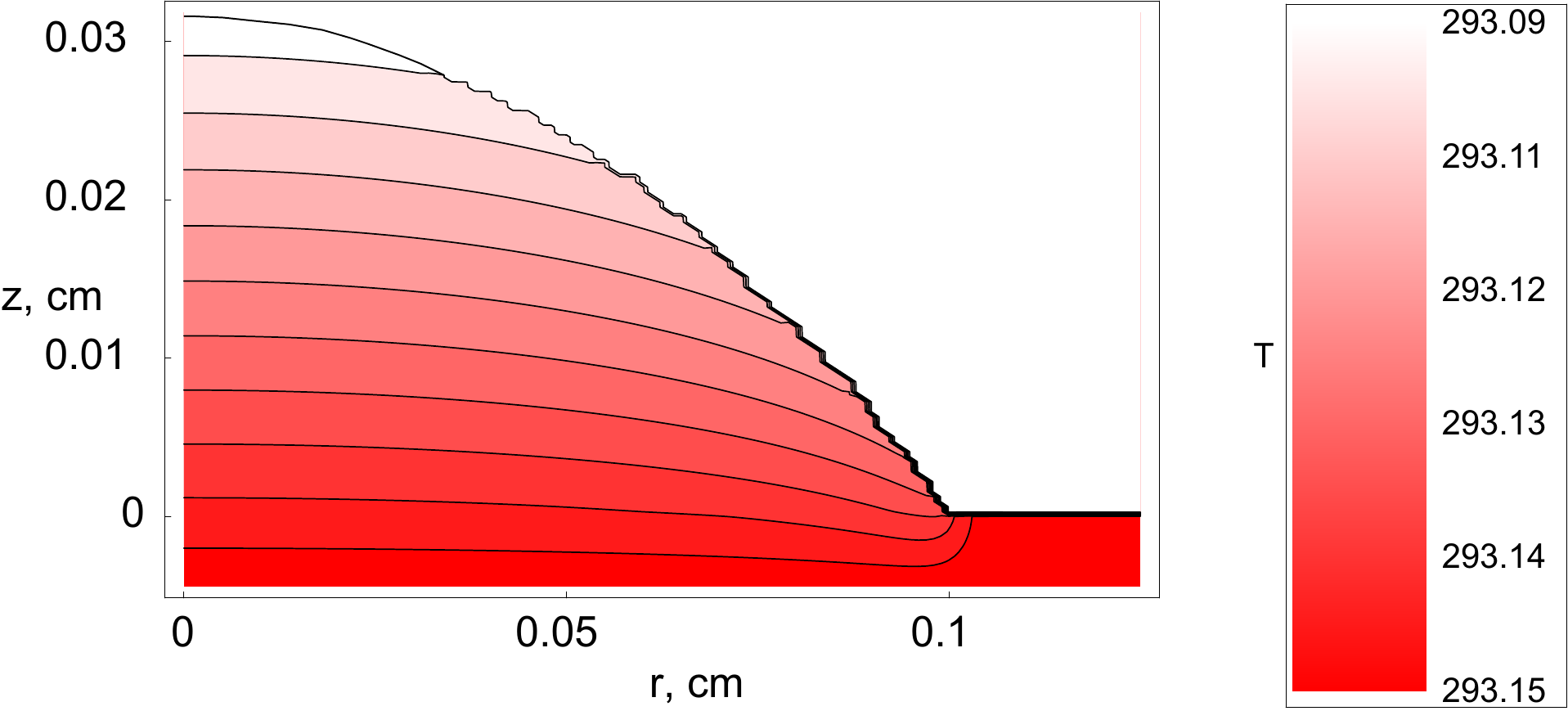}}
\hspace*{-0.2cm}
\raisebox{0.7cm}{\includegraphics[width=0.48\textwidth]{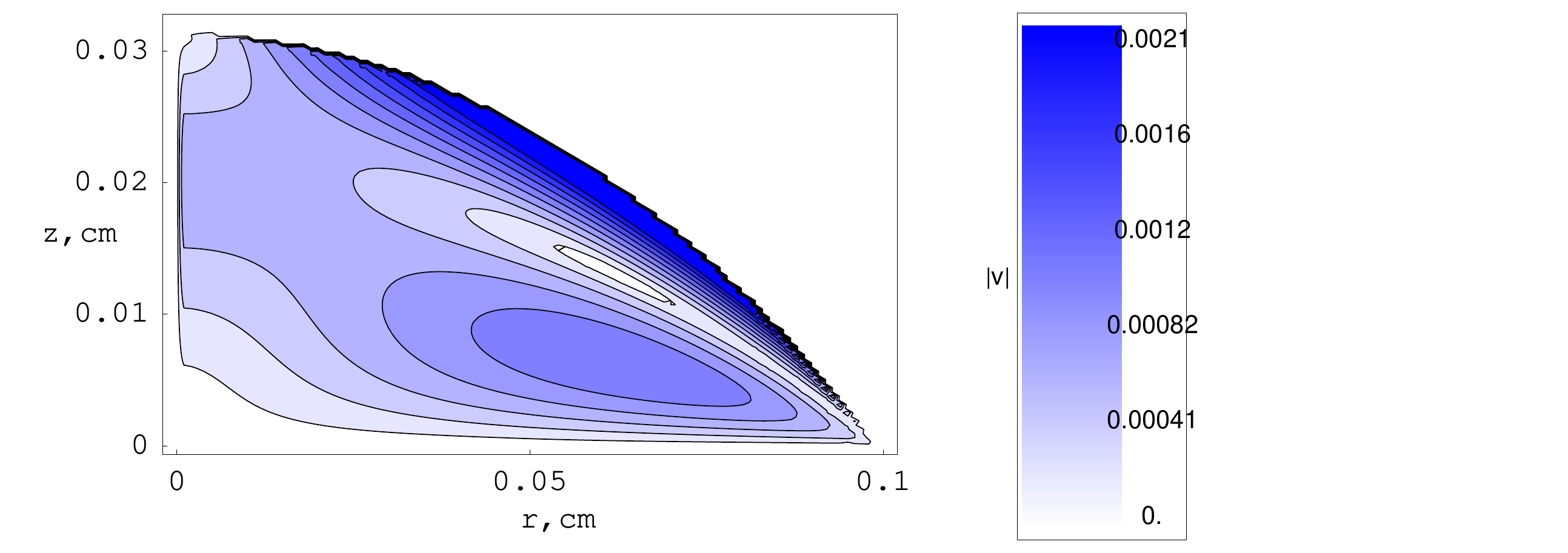}}
\hspace*{-1.9cm}
\includegraphics[width=0.25\textwidth]{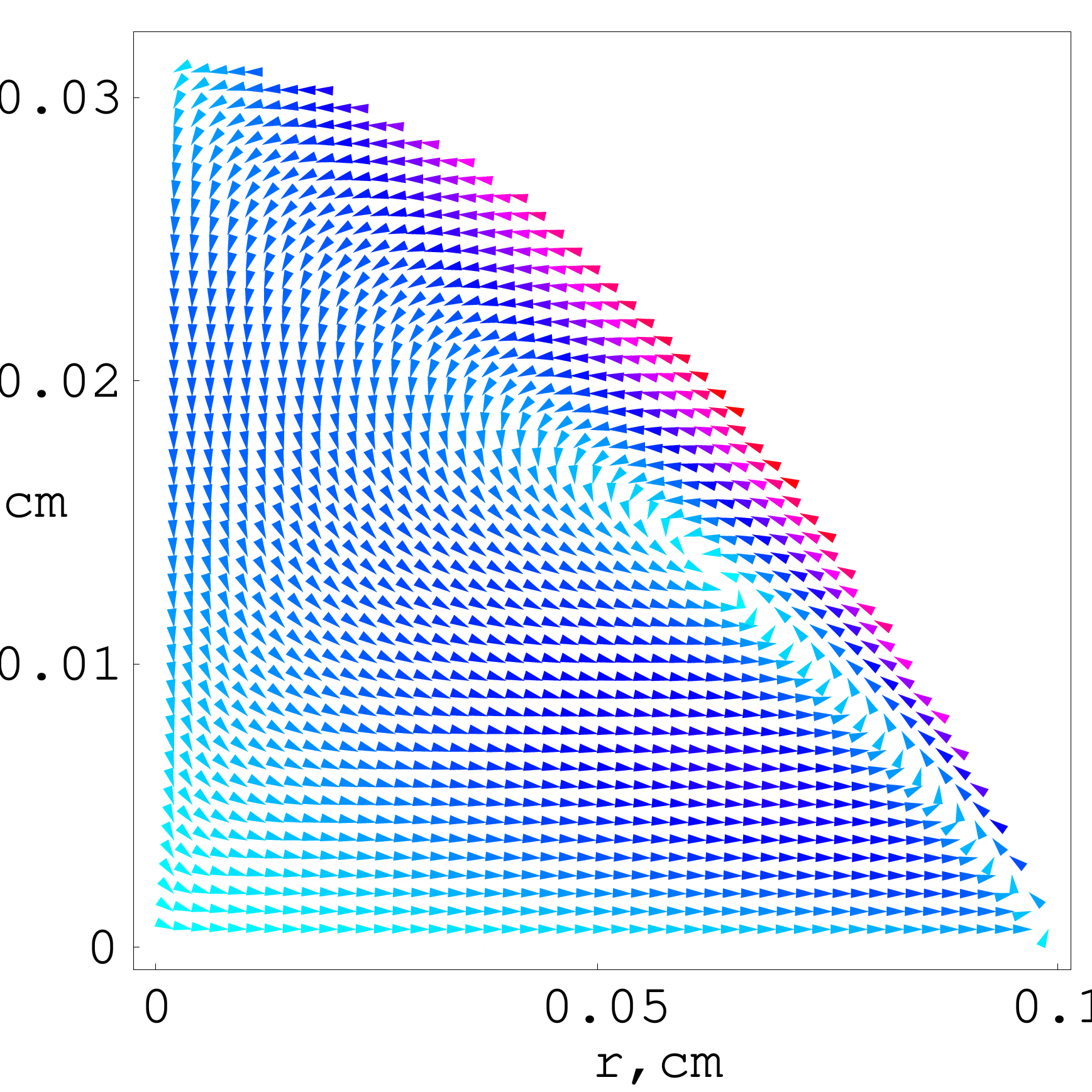}
\label{drop1}
\end{figure}
\begin{figure}[th]
\caption{Temperature distribution inside the droplet and the substrate correspondingly,
distribution of absolute value of velocity and vector field plot of velocity
in a reversed single-vortex regime obtained for 1-hexanol, $\theta=35^\circ$, $k_R=0.01$, $h_R=h_S/R=0.05$.}
\includegraphics[width=0.55\textwidth]{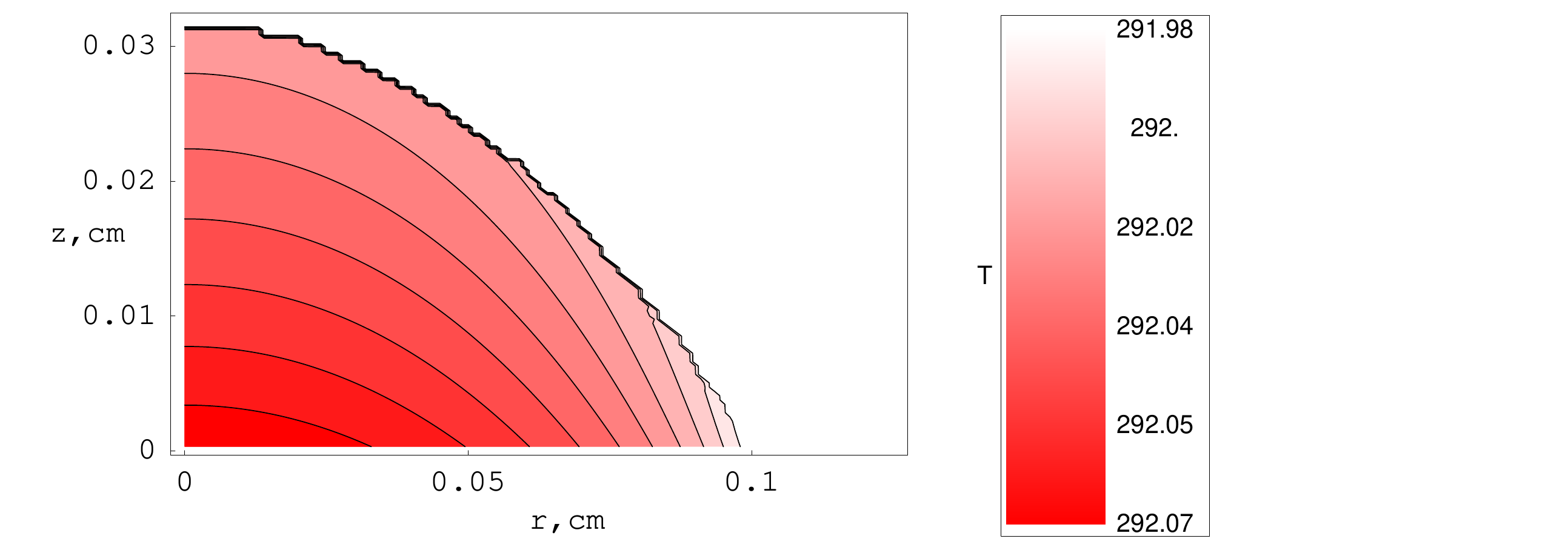}
\hspace*{-2.5cm}
\includegraphics[width=0.55\textwidth]{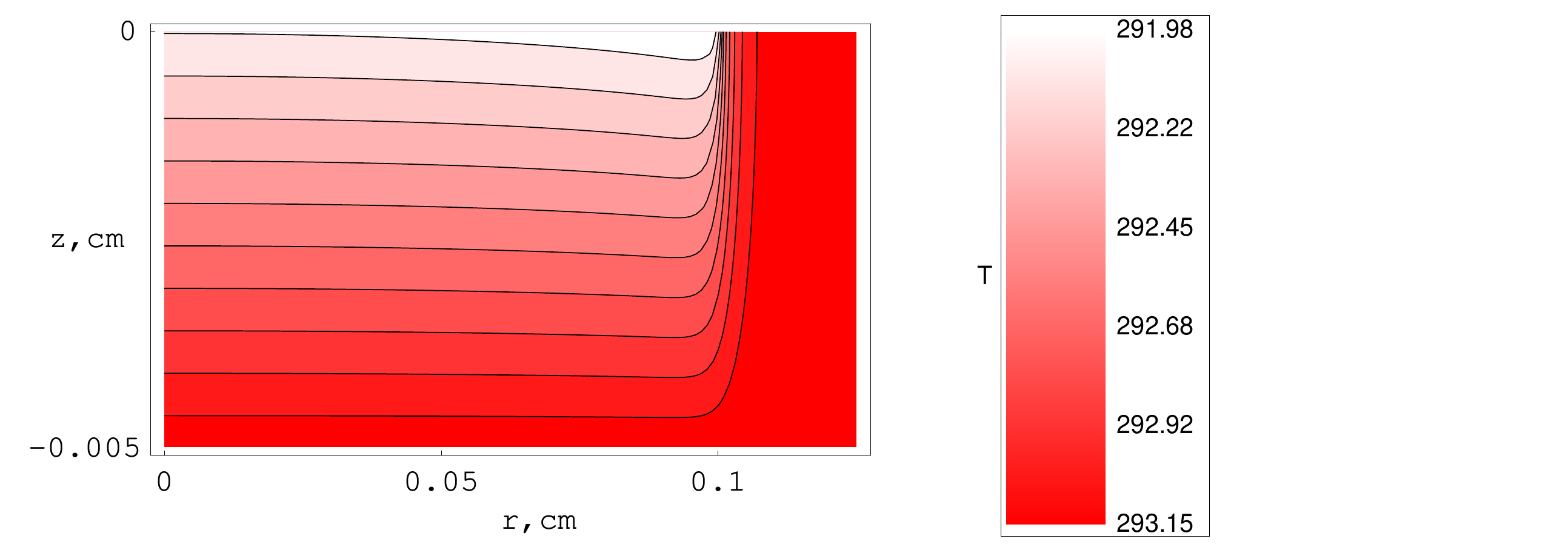}
\hspace*{-1.5cm}
\raisebox{0.7cm}{\includegraphics[width=0.68\textwidth]{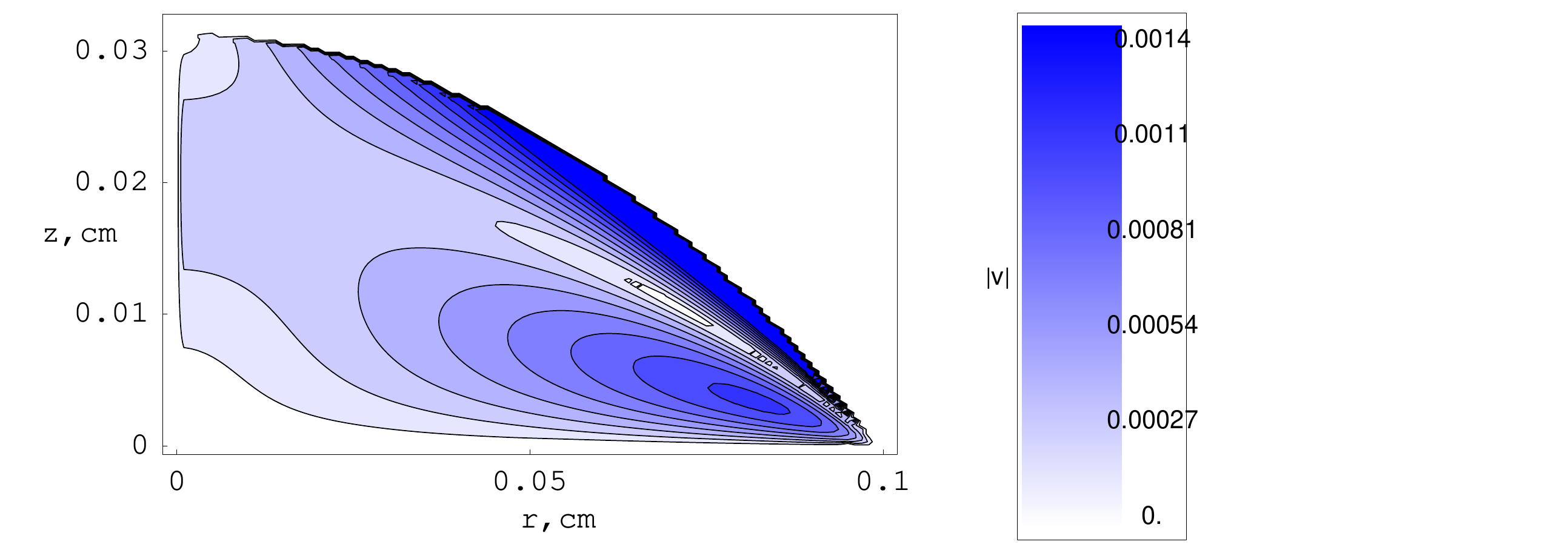}}
\hspace*{-2cm}
\includegraphics[width=0.3\textwidth]{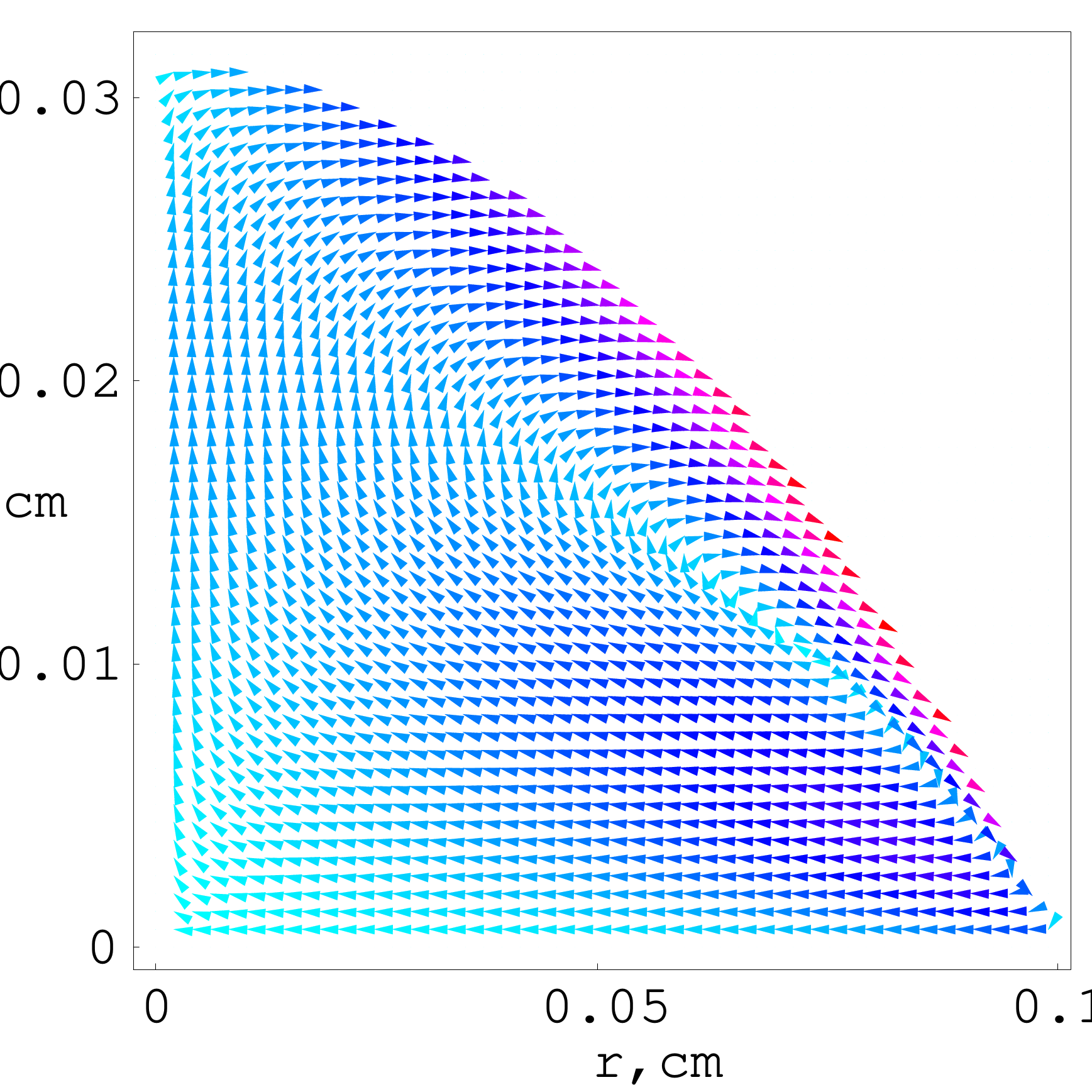}
\label{dropR}
\end{figure}
\begin{figure}[th]
\caption{Temperature distribution inside the droplet and the substrate correspondingly,
distribution of absolute value of velocity and vector field plot of velocity
for a droplet of 1-hexanol in a two-vortex
regime obtained for $\theta=22.44^\circ$, $k_R=0.2$, $h_R=h_S/R=0.5$.}
\includegraphics[width=0.45\textwidth]{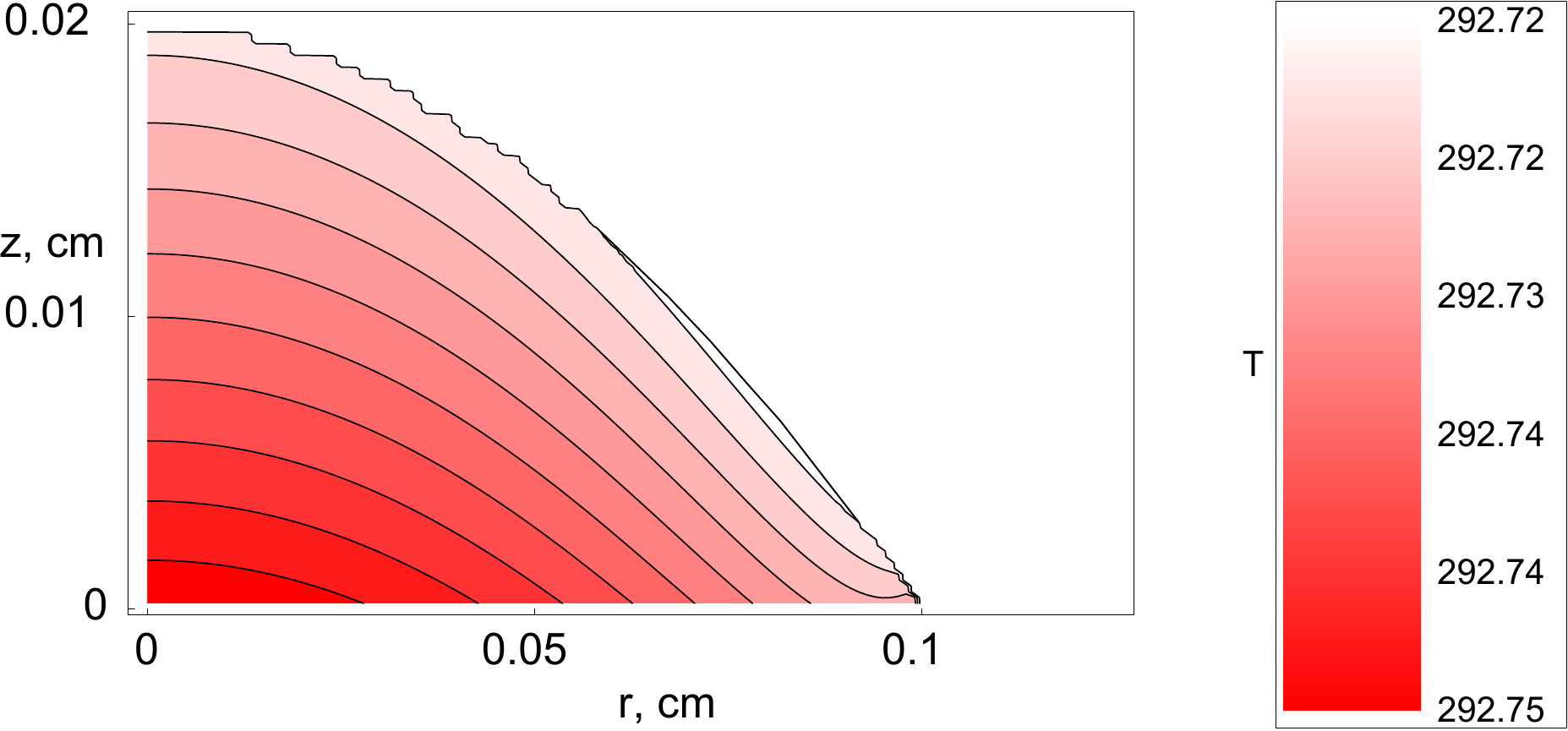}
\includegraphics[width=0.53\textwidth]{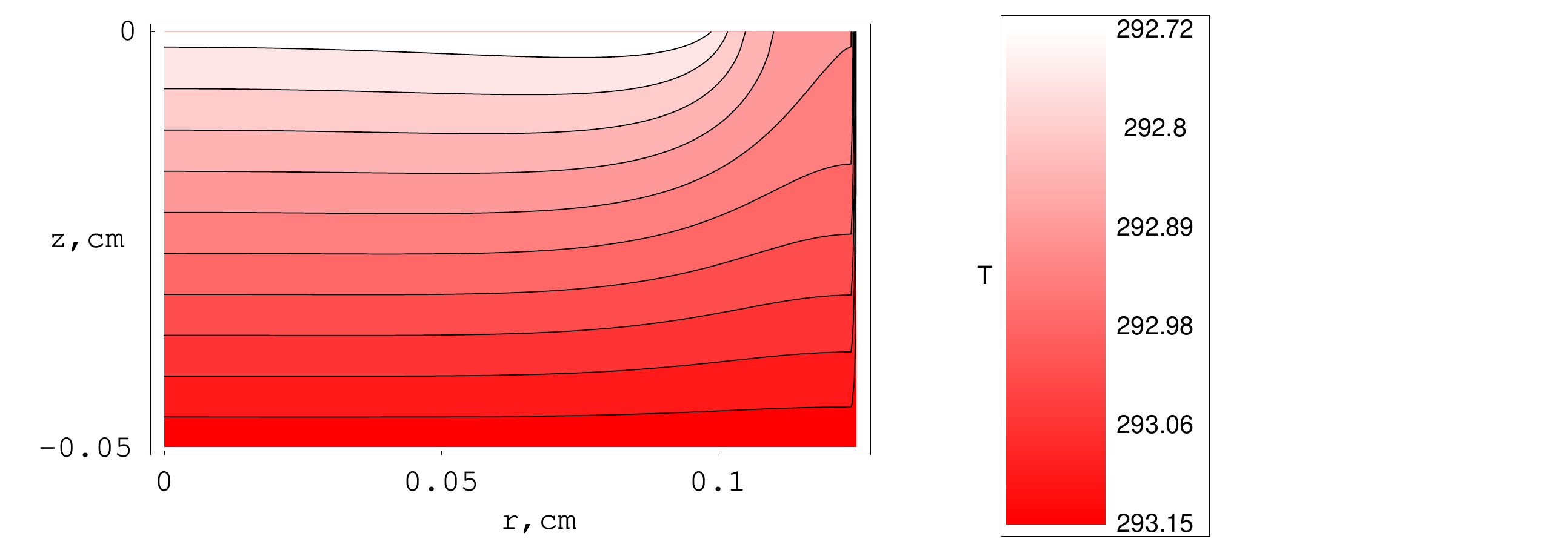}
\hspace*{-1.0cm}
\raisebox{0.7cm}{\includegraphics[width=0.68\textwidth]{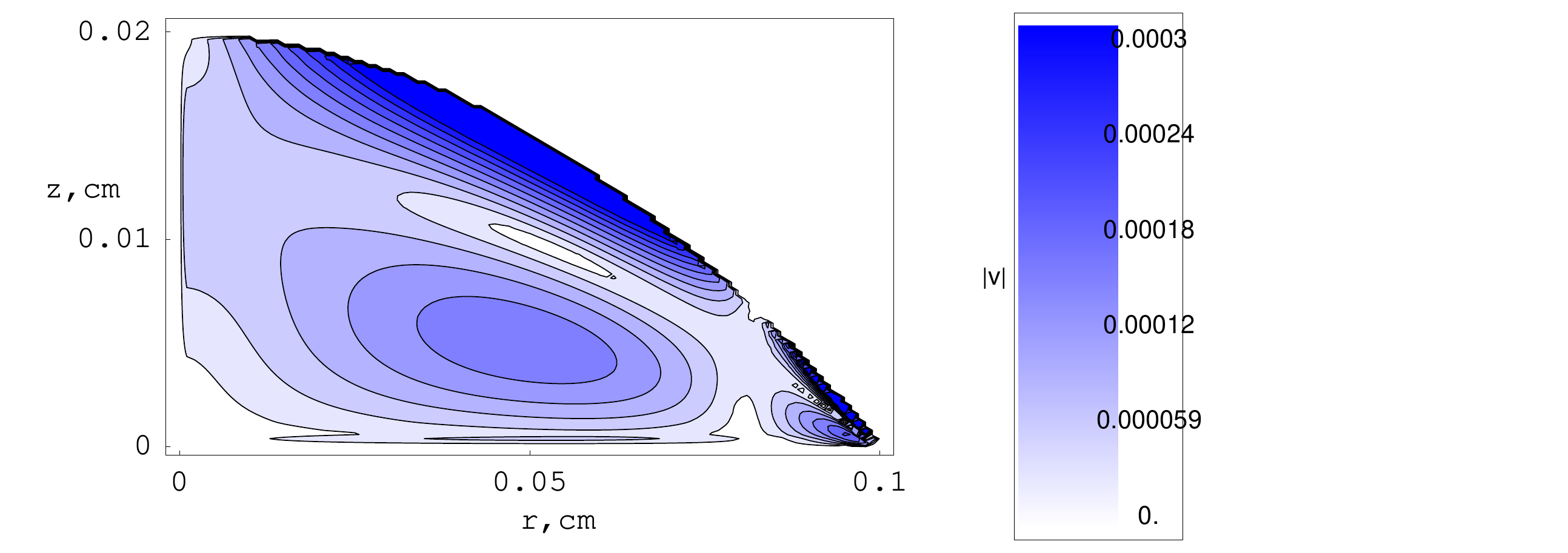}}
\hspace*{-2cm}
\includegraphics[width=0.3\textwidth]{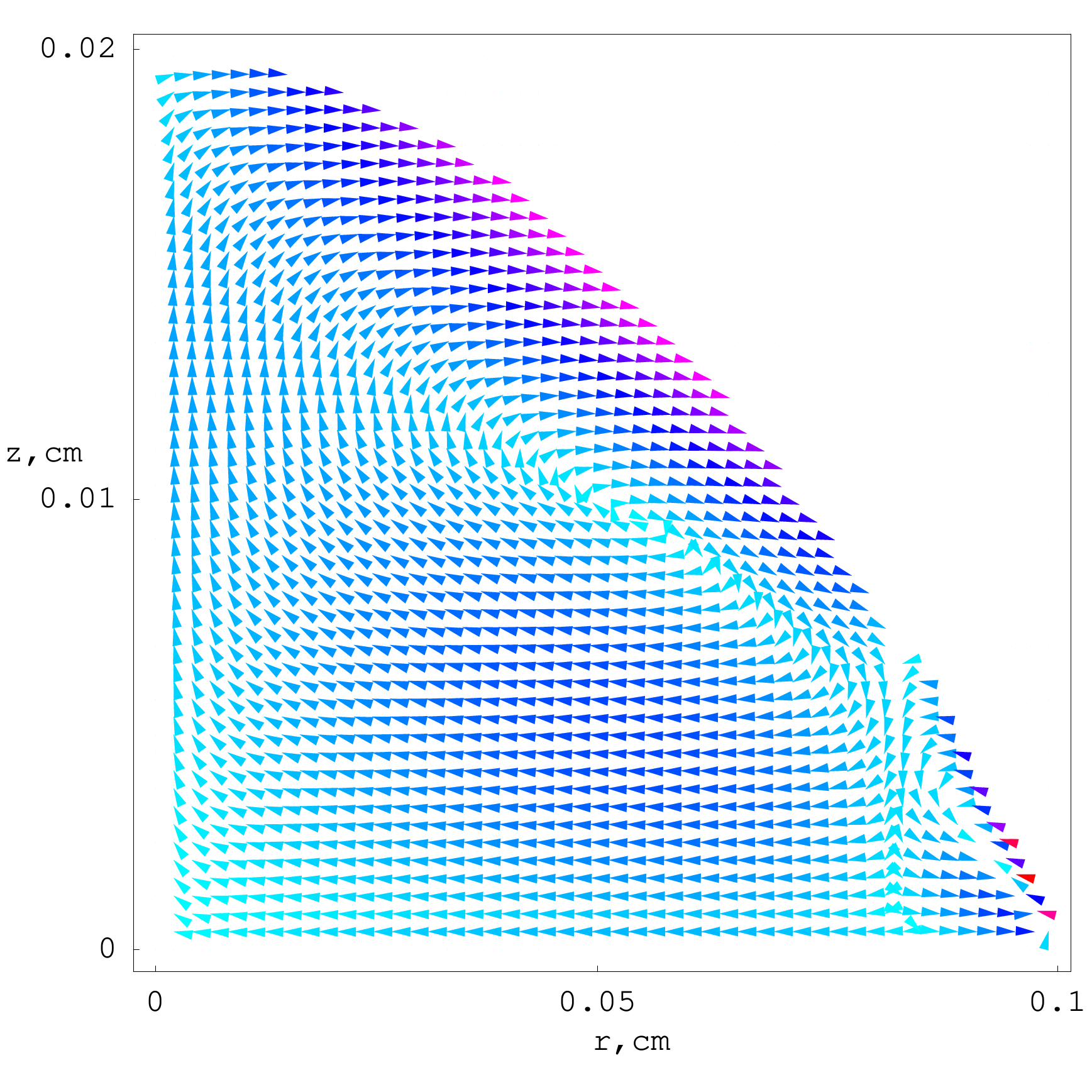}
\label{drop2}
\end{figure}
\begin{figure}[th]
\caption{Temperature distribution inside the droplet and the substrate correspondingly,
distribution of absolute value of velocity and vector field plot of velocity
for a droplet of 1-hexanol in a three-vortex
regime obtained for $\theta=31.22^\circ$, $k_R=0.2$, $h_R=h_S/R=0.1$.}
\includegraphics[width=0.45\textwidth]{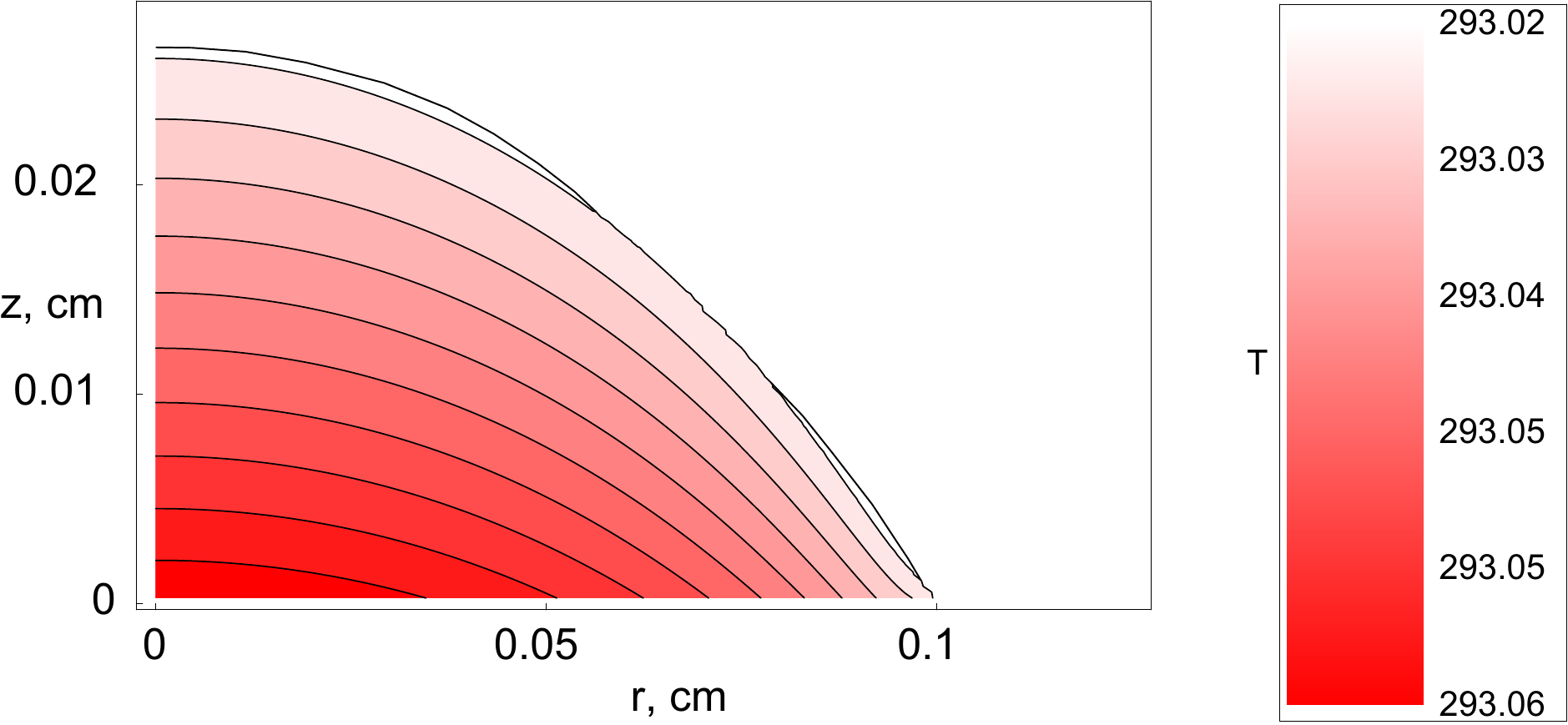}
\includegraphics[width=0.53\textwidth]{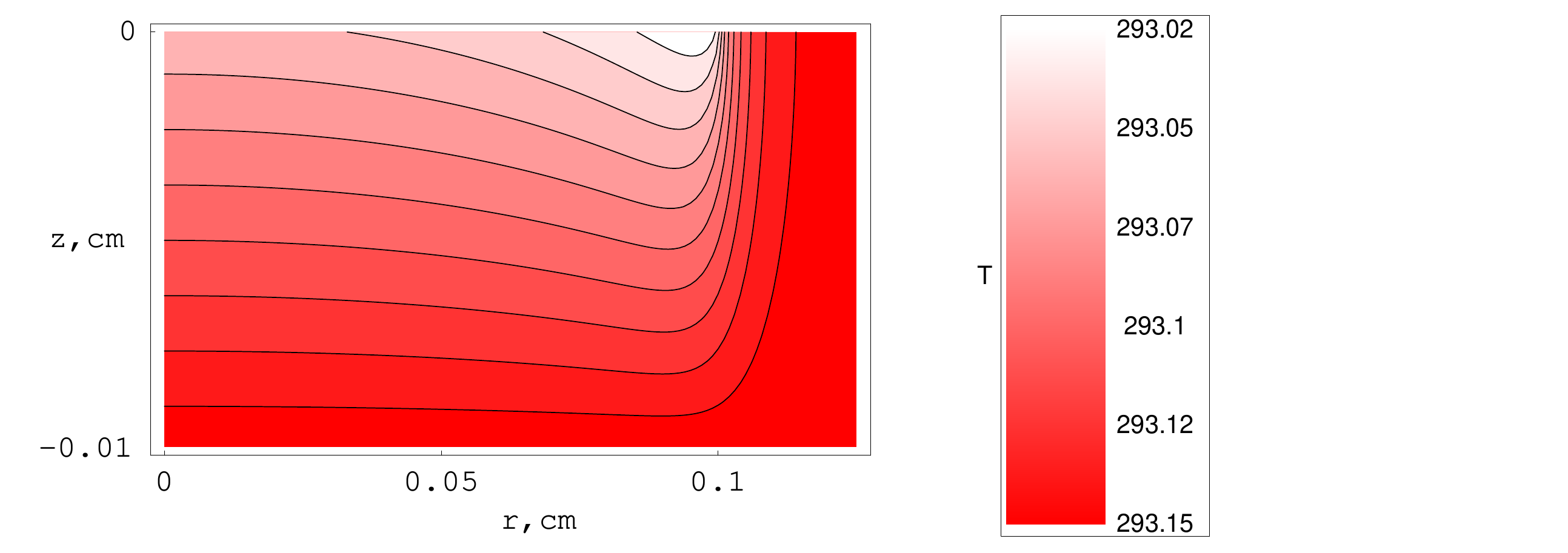}
\hspace*{-1.0cm}
\raisebox{0.7cm}{\includegraphics[width=0.68\textwidth]{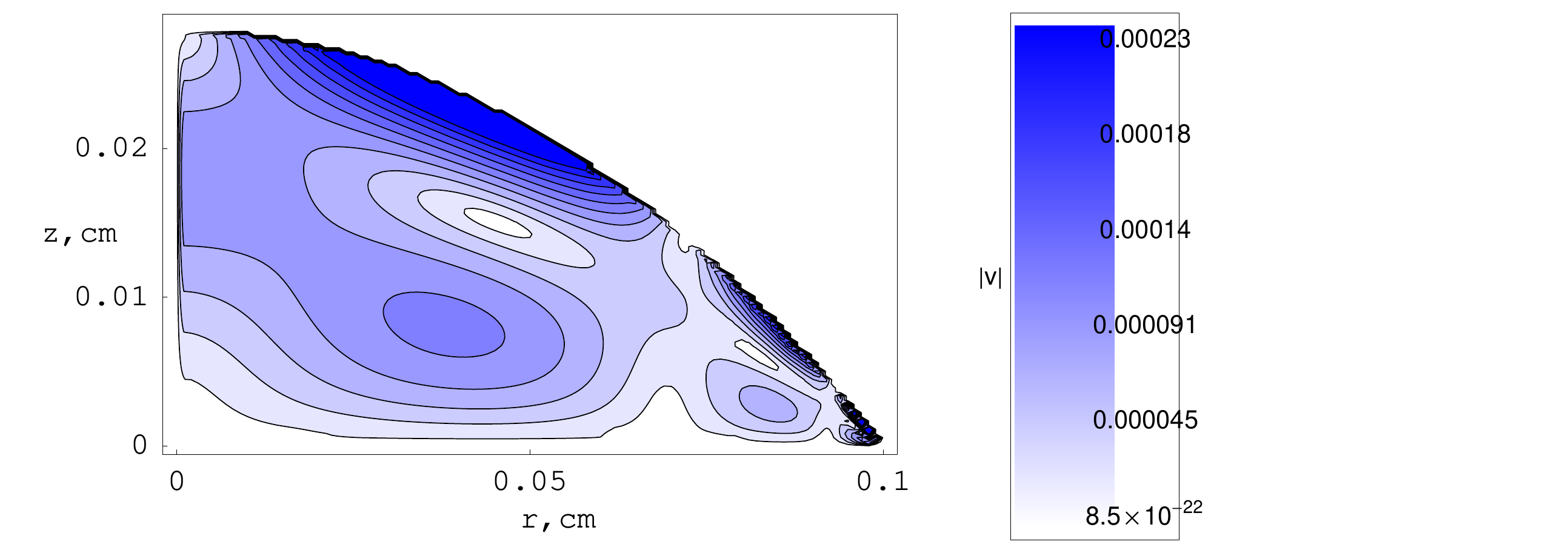}}
\hspace*{-2cm}
\includegraphics[width=0.3\textwidth]{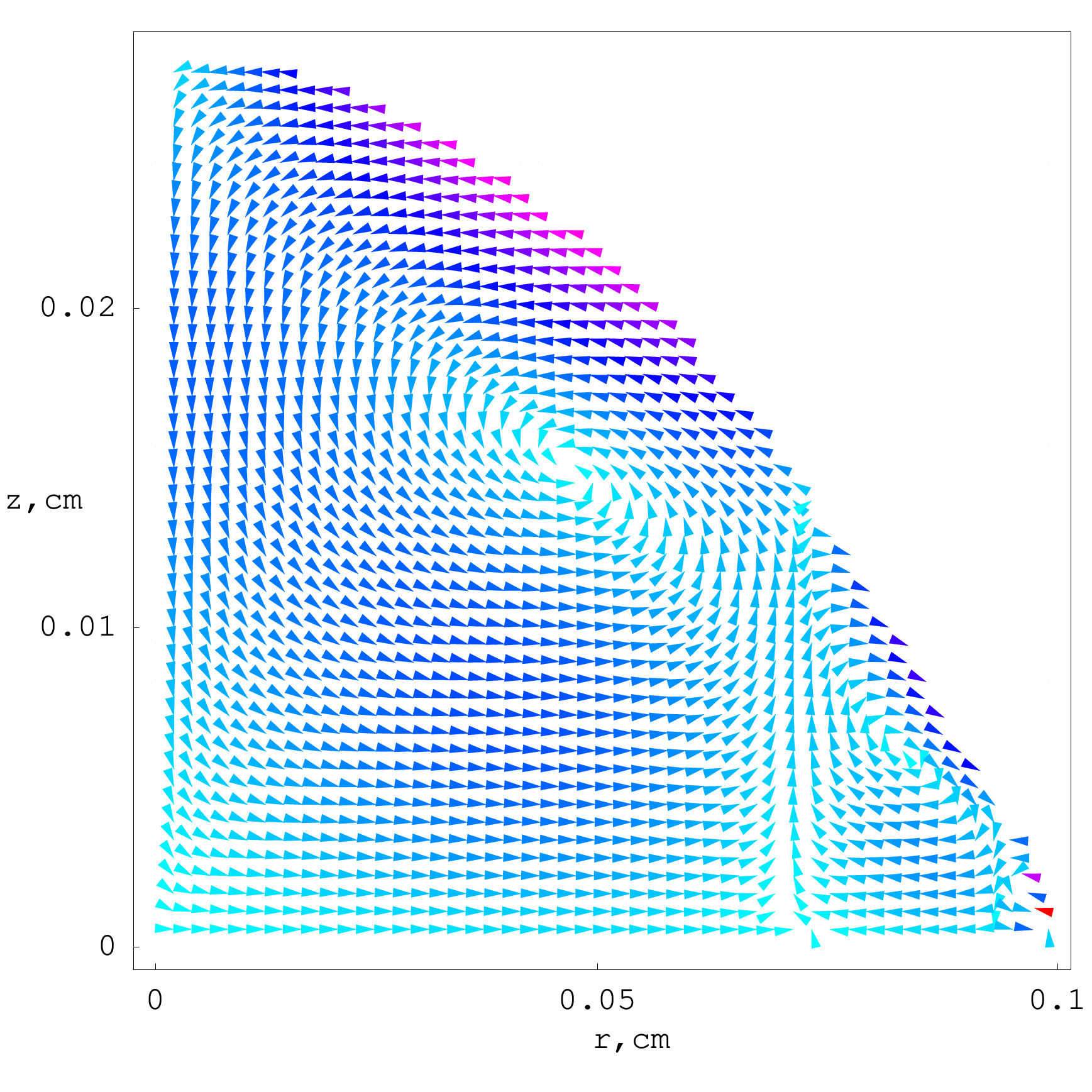}
\label{drop3}
\end{figure}
\begin{figure}[th]
\caption{Surface temperature distribution for a single-vortex, reversed single-vortex,
two-vortex and three-vortex regimes corresponding to the 1-hexanol droplets in Figs.~\ref{drop1}-\ref{drop3}.}
\label{TsurfPlots}
\includegraphics[width=0.245\textwidth]{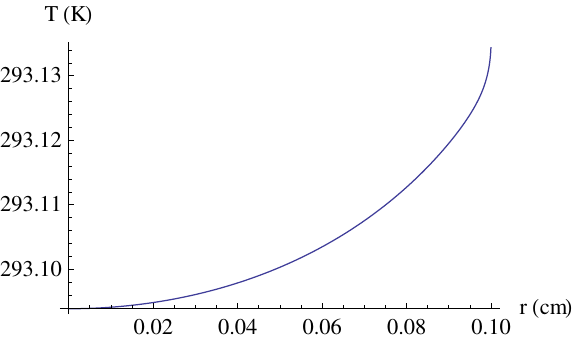}
\includegraphics[width=0.245\textwidth]{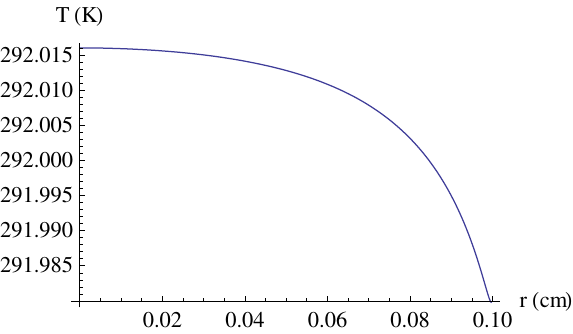}
\includegraphics[width=0.245\textwidth]{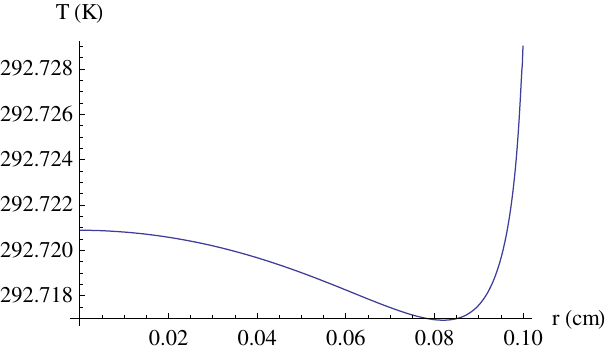}
\includegraphics[width=0.245\textwidth]{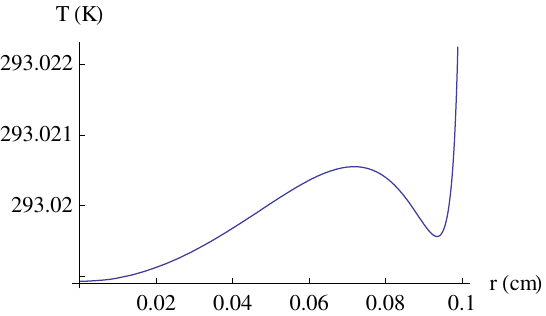}
\end{figure}

In the second part of the method one takes
$z$-derivative implicitly and represents Eqn.~(\ref{EqThermal}) as
\be
\frac{T_{ij}^{n+1}-T_{ij}^{n+1/2}}{\kappa h_t/2}=
\hat\delta_r^2 T_{ij}^{n+1/2}+\hat\delta_z^2 T_{ij}^{n+1}+
\frac{1}{r}\frac{T_{i+1,j}^{n+1/2}-T_{i-1,j}^{n+1/2}}{a+b}.
\ee
For given temperature at time step $n+1/2$ this
expression takes the form
\be
c_j''T_{i,j-1}^{n+1}+(d_j''+2/(\kappa h_t))T_{ij}^{n+1}+e_j''T_{i,j+1}^{n+1}=
c_i'T_{i-1,j}^{n+1/2}+(d_i'+2/(\kappa h_t))T_{ij}^{n+1/2}+e_i'T_{i+1,j}^{n+1/2},
\label{implicit2}
\ee
where the coefficients are given in (\ref{koeff1})-(\ref{koeff4}).
For each $i$ the tridiagonal matrix algorithm is used
to solve the set of equations (\ref{implicit2}) for the temperature at time step $n+1$.
The boundary condition $T^{n+1}_{S,i0}=T_0$, the boundary interpolation
at the droplet surface (see below) and the matching condition
$T_{i0}=c_ST_{i1}+c_LT_{S,i,n-1}$ are solved together with the 
set of equations (\ref{implicit2}) by the tridiagonal matrix algorithm, 
where $c_S=k_Lh_{yS}/(k_Lh_{yS}+k_Sh_{yL})$, $c_L=1-c_S$.
For $i>n$ the boundary condition at the substrate--gas interface is also used.

At the droplet surface we use the boundary interpolation.
Consider a mesh point $D(i,j)$ close to the surface, where $j>1$.
The point $D$ is inside the drop and at least
one of its nearest neighbors is outside the drop.
In linear approximation $T(r,z)=a+br+cz$
in a vicinity of the point $D(i,j)$.
We denote the temperatures at points $D(i,j)$,
$B(i-1,j)$ and $C(i,j-1)$ as $T_D$, $T_B$ and $T_C$
correspondingly; also let $G=\p T/\p n=-LJ(r)/k$
at a point $A$ of the drop surface near $D$.
Then one has
\begin{equation}
b\sin\phi+c\cos\phi = G,\qquad
a+b(i-1)h_r+cjh_z = T_B,\qquad
a+bih_r+c(j-1)h_z = T_C.
\end{equation}
The solution of the set of equations is
\begin{eqnarray}
\label{ukoeff1}
a&=&\left(
h_r\cos\phi(iT_B-(i-1)T_C)+h_z\sin\phi(T_B+j(T_C-T_B))
-h_rh_zG(j+i-1) \right)/R\\
b&=&(h_zG+(T_C-T_B)\cos\phi)/R\\
c&=&(h_rG+(T_B-T_C)\sin\phi)/R\\
R&=&h_r\cos\phi+h_z\sin\phi.
\label{ukoeff4}
\end{eqnarray}
In the first part of the alternating direction method
the calculations of rows proceed towards larger values
of $j$. For this reason one should consider here
$T_C$ and $G$ as given quantities, whereas
$T_B$ and $T_D$ are unknown. It is convenient under
these conditions to represent (\ref{ukoeff1})-(\ref{ukoeff4})
as
\begin{equation}
a=a_0+a_1T_B,\qquad
b=b_0+b_1T_B,\qquad
c=c_0+c_1T_B,
\end{equation}
and obtain explicit expressions for $a_0,a_1,b_0,b_1,c_0,c_1$.
This results in linear relation between $T_D$ and $T_B$
\begin{equation}
T_D=a+br_D+cz_D=(a_0+b_0r_D+c_0z_D)+(a_1+b_1r_D+c_1z_D)T_B,
\end{equation}
which can be transformed to the form
$d_{i-1}'T_{i-1,j}^{n+1/2}+e_{i-1}'T_{ij}^{n+1/2}=b_{i-1}'$.
This completes the set of equations (\ref{implicit1}) for the
tridiagonal matrix algorithm.
Here $d_{i-1}'=-(a_1+b_1r_D+c_1z_D)$, $e_{i-1}'=1$,
$b_{i-1}'=a_0+b_0r_D+c_0z_D$.

To carry out the boundary interpolation
in the second part of the alternating direction
method, similar expressions can be derived to relate
$T_D$ and $T_C$.
A similar boundary interpolation method is also derived 
for $j\leq 1$.

\section{Results and discussions}

\begin{figure}[tbh]
\caption{Results for $\theta$ vs $k_R=k_S/k_L$, where 
a) $h_R=0.01$, b) $h_R=0.05$, c) $h_R=0.1$, d) $h_R=0.2$, e) $h_R=0.5$, f) $h_R=1$.}
\includegraphics[width=0.32\textwidth]{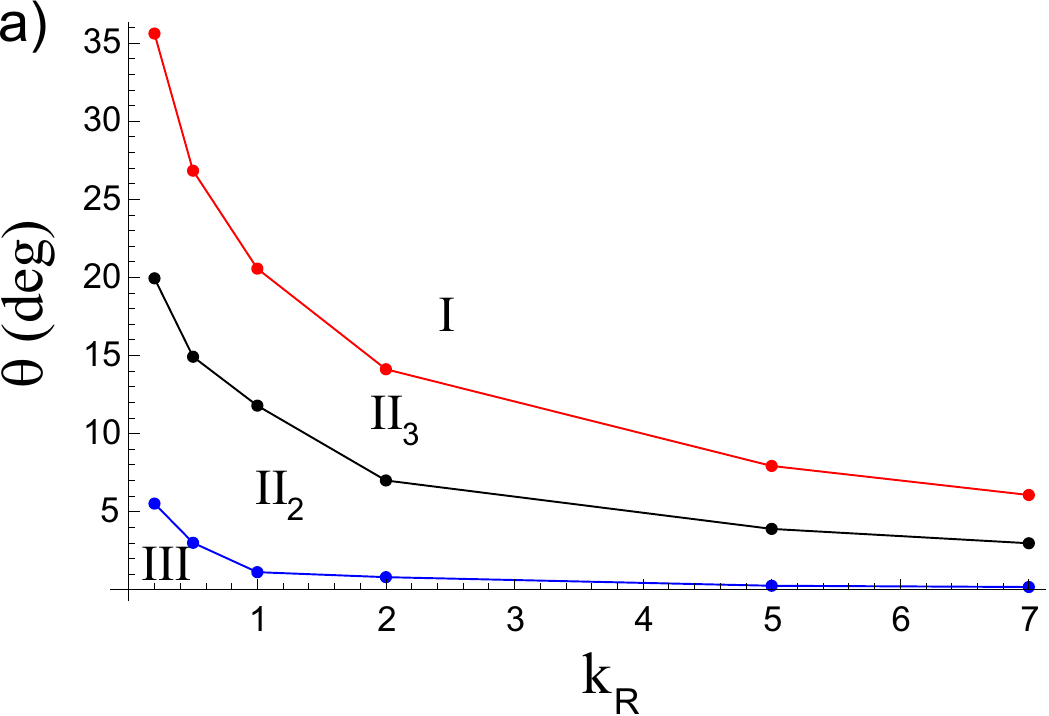}
\includegraphics[width=0.32\textwidth]{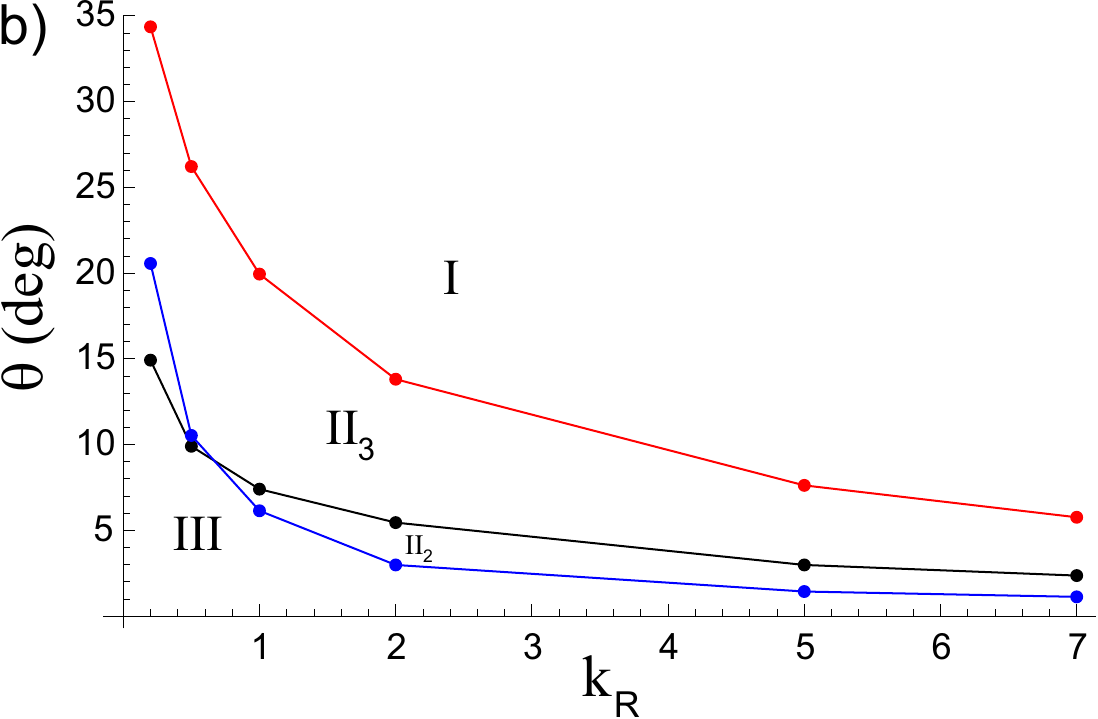}
\includegraphics[width=0.32\textwidth]{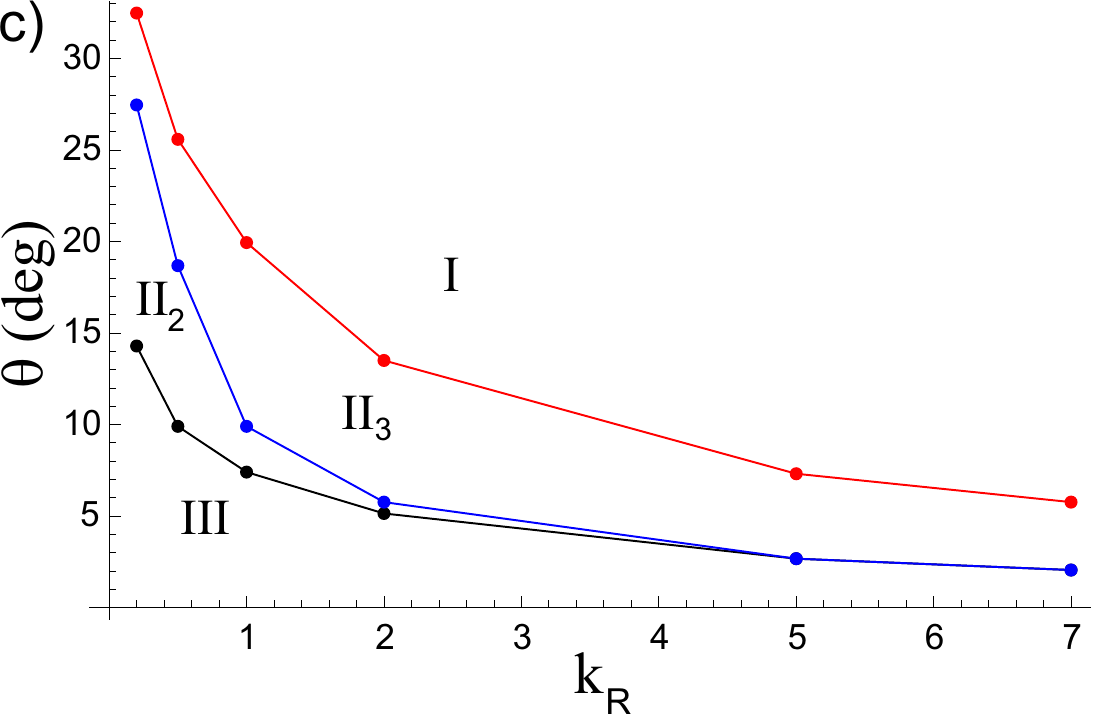}
\includegraphics[width=0.32\textwidth]{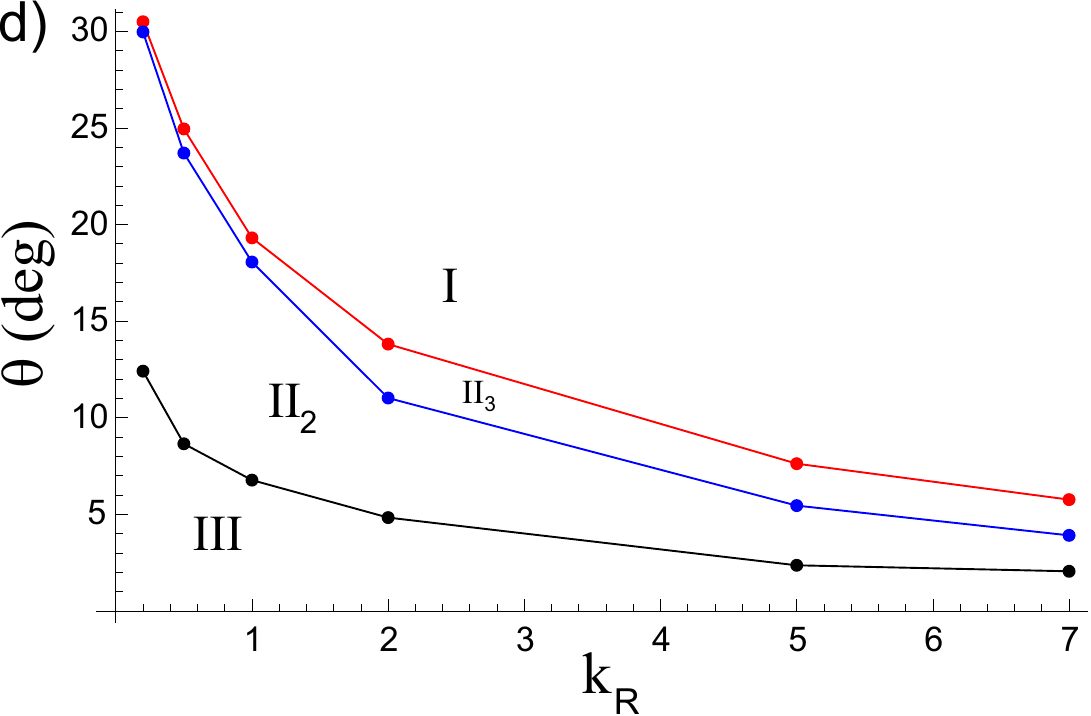}
\includegraphics[width=0.32\textwidth]{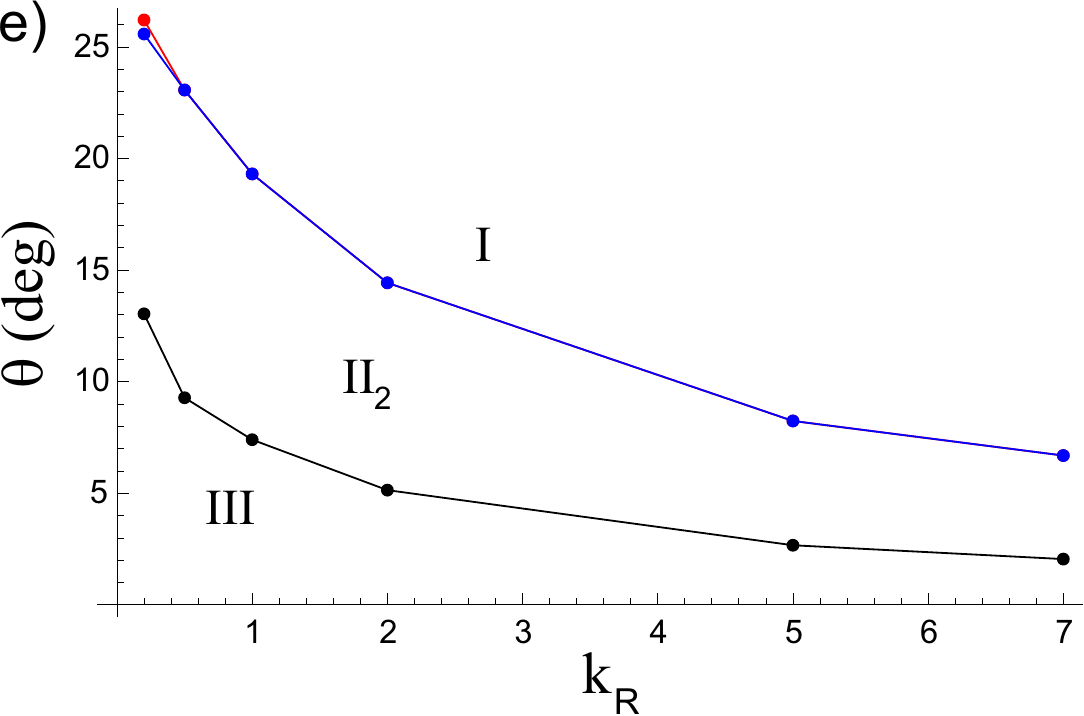}
\includegraphics[width=0.32\textwidth]{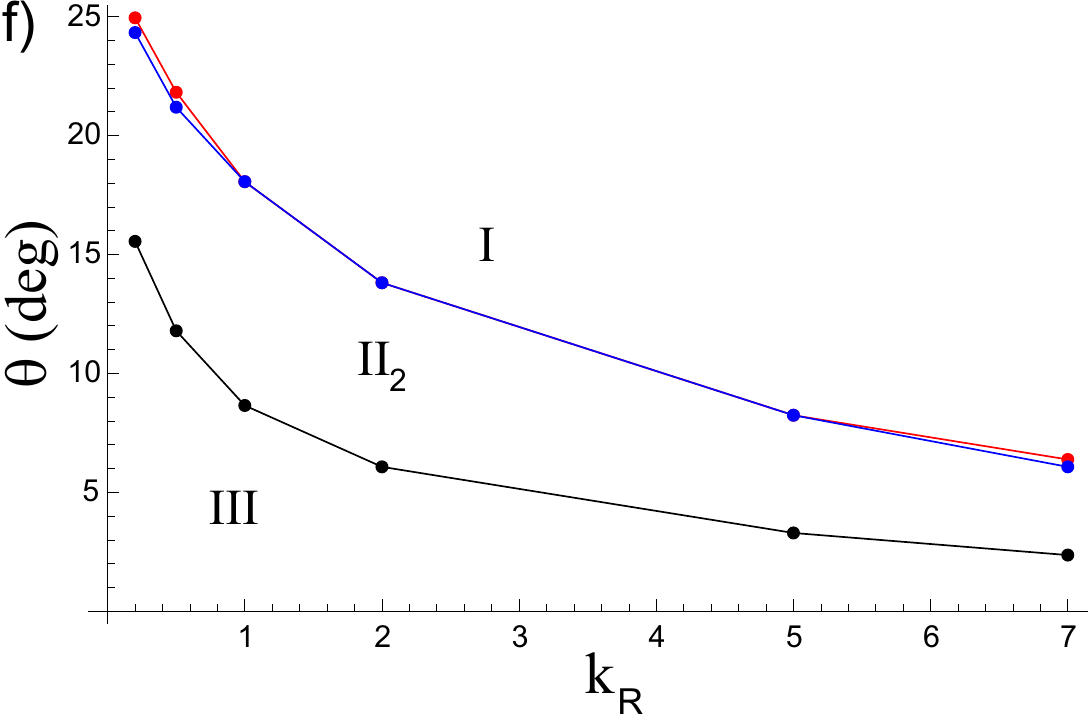}
\label{diag}
\end{figure}
\begin{figure}[tbh]
\caption{Dependence of $\theta_{crit}$ on $h_R=h_S/R$ for
a) the transition corresponding to the change of sign of tangential
component of the temperature gradient at the
liquid--vapor interface near the contact line;
b) the transition to a monotonically increasing surface temperature dependence on $r$;
c) the transition corresponding to the change of sign of tangential component of
the temperature gradient at the
liquid--vapor interface near the droplet apex.}
\includegraphics[width=0.30\textwidth]{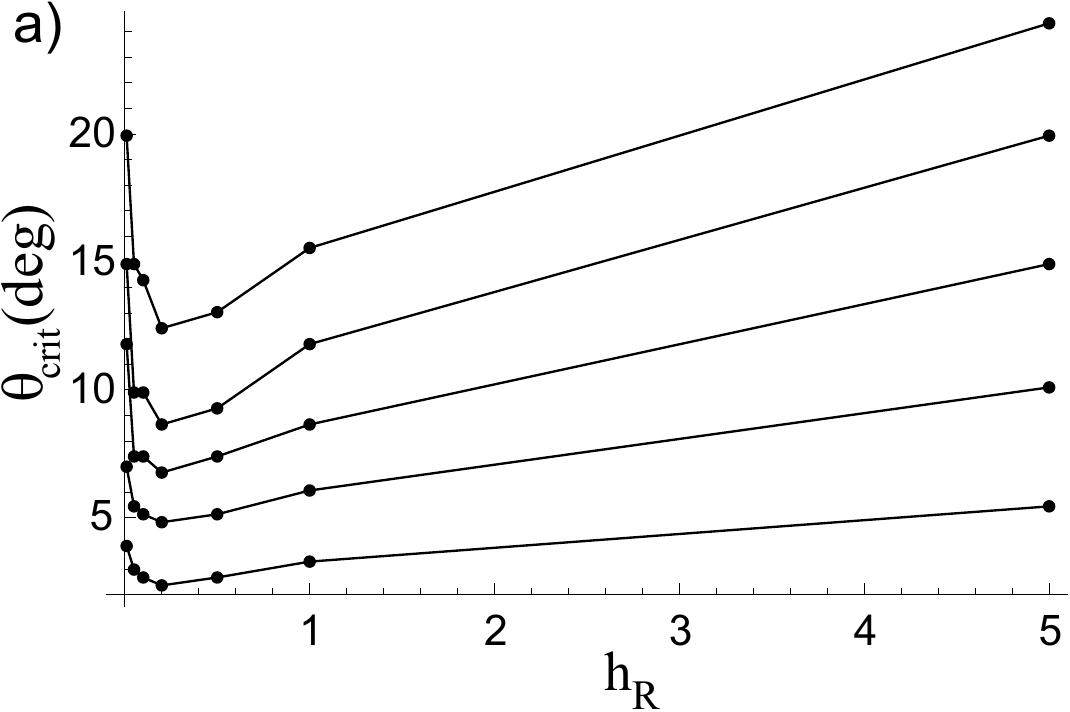}
\includegraphics[width=0.34\textwidth]{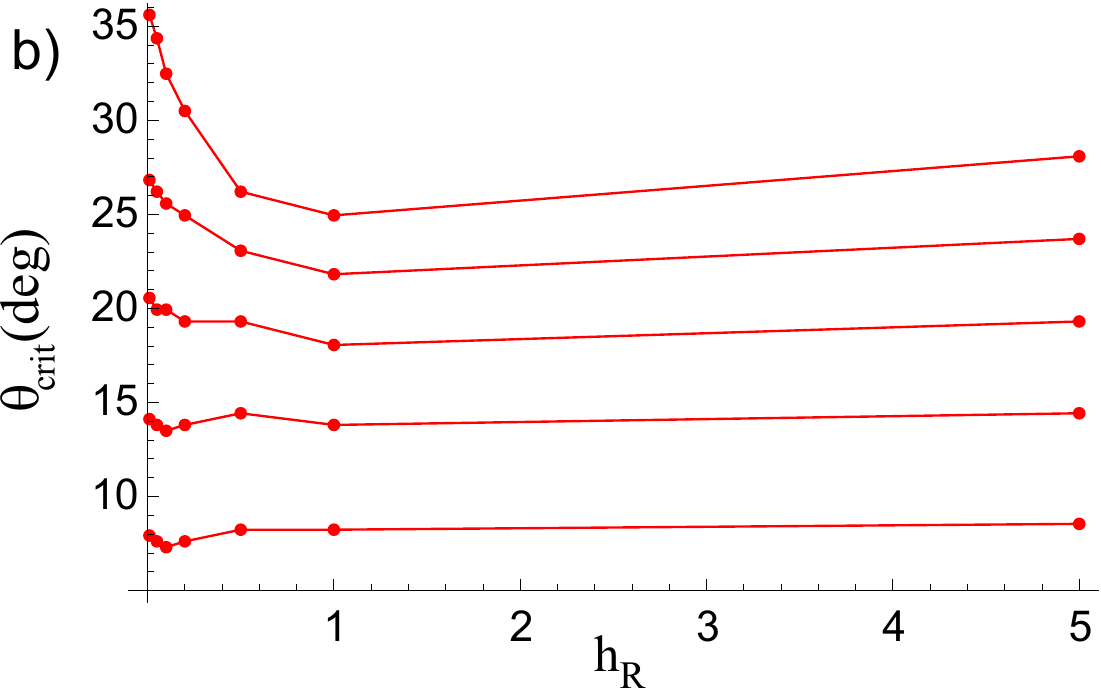}
\includegraphics[width=0.34\textwidth]{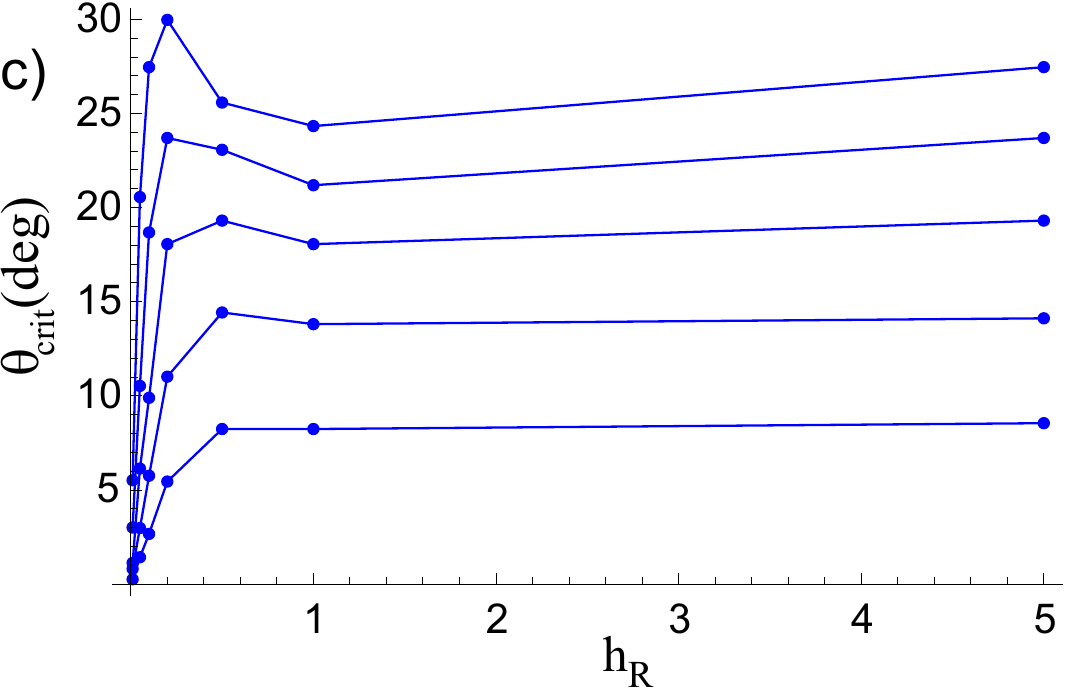}
\label{plotABF}
\end{figure}

We have carried out the numerical simulation of the evaporation and fluid dynamics of
droplets of capillary size and studied the dependence of the results 
on the characteristics of the substrate. 
The thermal conduction equation and the Navier--Stokes equations have been solved using
the numerical method described in Section~\ref{MethodsSec},
where the equations and the boundary conditions used are presented as well.

Figs.~\ref{drop1}--\ref{TsurfPlots} show four different regimes
of fluid flow structure in an evaporating sessile droplet.
The parameter set used is taken for 1-hexanol and presented in Table~\ref{ParamTable},
where the notations are also introduced.
The single-vortex regime is shown in Fig.~\ref{drop1},
the reversed single-vortex regime is shown in Fig.~\ref{dropR},
the two-vortex regime is shown in Fig.~\ref{drop2},
and the three-vortex regime is shown in Fig.~\ref{drop3}.
Figs.~\ref{drop1}--\ref{drop3} demonstrate spatial temperature 
distributions both inside the droplet and the substrate, 
distributions of absolute value of the fluid velocity and
vector field plots of the velocity.
The temperature distributions inside the droplet and the substrate
are shown separately in Figs.~\ref{dropR}--\ref{drop3}
because different temperature scales are used here for the fluid and for the substrate.
Using both plot of the velocity absolute value and vector field plot of the velocity
allow to give a thorough picture of the velocity distribution.
The basic difference between the four regimes is summarized in
Fig.~\ref{TsurfPlots}, which shows the surface temperature
distribution corresponding to the droplets in Figs.~\ref{drop1}--\ref{drop3}.

Since 1-hexanol is weakly volatile liquid, the evaporation process
is comparatively slow, and the corresponding P\'eclet number 
$Pe=\overline{u}R/\kappa$ is much smaller than unity for
the droplets of relatively small size considered (here $\overline{u}$
is a characteristic value of fluid velocity).
The assumption $Pe\ll 1$ is justified by the numerical
results for the velocity values shown in the absolute value plots 
in Figs.~\ref{drop1}--\ref{drop3}. Hence, the 
convective heat transfer is negligibly small. Also we consider the 
evaporation as a quasistationary process, which is realized when the
transient time for heat transfer $t_{\text{heat}}=R(h_0+h_S)/\kappa$, 
transient time for momentum transfer $t_{\text{mom}}=\rho Rh_0/\eta$ and
transient time for vapor phase mass transfer $t_{\text{mass}}=
\rho_{\text{vap}}/\rho\cdot t_f$ are much smaller than the total drying 
time $t_{\text{f}}\approx 0.2\rho R h_0/(Du_{\text{s}})$. 
Here $h_0$ is the droplet height and $h_S$ is the substrate thickness.
In addition, the droplet shape can be described with good accuracy within the 
spherical cap approximation when the capillary number $Ca=\eta
\overline{u}/\sigma$ and the Bond number $Bo=\rho g h_0 R/(2\sigma
\sin\theta)$ are much smaller than unity~\cite{Barash2009}. 

Since driving forces for fluid convection are the Marangoni forces
and $\p\sigma/\p T<0$, the fluid usually flows from a surface region with higher
temperature to that with lower temperature.
Therefore, the single-vortex flow corresponds to 
a surface temperature, which monotonically increases with increasing
the axial coordinate $r$ (see Figs.~\ref{drop1},\ref{TsurfPlots}).
The reversed single-vortex flow (see Figs.~\ref{dropR},\ref{TsurfPlots}) corresponds to
a monotonically decreasing surface temperature dependence on $r$.
The surface temperature, which induces a two-vortex or three-vortex flow,
is found to be a nonmonotonic function of $r$ with a single extremum 
(in addition to extrema at the apex and at the contact line)
in the two-vortex case and with both maximal and minimal values 
in the three-vortex case (see Figs.~\ref{drop2}--\ref{TsurfPlots}).
Fig.~\ref{TsurfPlots} shows surface temperature distributions
for the droplets in Figs.~\ref{drop1}--\ref{drop3}.
We do not aim at listing all possible types of surface temperature distribution in Fig.~\ref{TsurfPlots}.
Also commonly observed is a two-vortex situation with a negative derivative at the edge,
positive derivative at the apex, and a sign change in slope elsewhere, which
is not shown in Fig.~\ref{TsurfPlots}.

The parameter regions of different types of surface temperature distribution
are shown in Fig.~\ref{diag} in the form of the ``phase 
diagram'' in the $k_R$--$\theta$ plane, taken for various
values of the substrate thickness. One can see that the region II 
of a nonmonotonic spatial dependence of the surface temperature consists
of two regions. 
One of them, II$_2$, identifies surface temperature distributions
with a single intermediate extremum, while another one, 
II$_3$, identifies surface temperature distributions
with both maximal and minimal values.
II$_2$ usually correspond to the two-vortex
flows, while II$_3$ corresponds to the flows with three vortices.
Black curve in Fig.~\ref{diag}
corresponds to the change of sign 
of the tangential component of temperature gradient 
at the liquid--vapor interface near the contact line.
Blue curve in Fig.~\ref{diag}
corresponds to the change of sign 
of the tangential component of temperature gradient 
at the liquid--vapor interface at the droplet apex.
Red curve in Fig.~\ref{diag}
corresponds to the transition between
a nonmonotonic dependence of surface temperature on $r$
and a monotonically increasing surface temperature dependence on $r$.

The ``phase diagram'' shows that the regime of convection 
changes during the evaporation process when the contact
angle diminishes.
For example, when $k_R$ is large, three vortices can appear
at relatively small contact angles (see also~\cite{Barash2009}).

The main features of a ``phase diagram'' can be qualitatively
understood as a result of matching the heat transfer 
through the solid--liquid interface and the 
heat flowing through the liquid--vapor interface,
which is associated with evaporative cooling.
When $k_R$ is small, the evaporative cooling is much stronger than the 
heat flow from the substrate. 
Since an inhomogeneous evaporation rate 
$J_s(r)=J_0(\theta)(1-r^2/R^2)^{-\lambda(\theta)}$
has its minimum at the apex, the surface region
near the apex should be warmer than at the contact line.
Thus, $k_R \ll 1$ results in a monotonic
decrease of the surface temperature with $r$, which
corresponds to the reversed single-vortex regime,
i.e., to the region III of the ``phase diagram''.
If, on the contrary, $k_R\gg 1$, then the
heat flowing through the solid--liquid interface is much stronger
than the heat flow due to evaporative cooling. Therefore, the area of liquid--vapor
interface close to the contact line is the warmest one
due to adjacent highly conducting substrate.
In this case the surface area close to the apex
is colder as compared to other parts of the droplet. 
Thus, $k_R\gg 1$  results in a monotonic
increase of the surface temperature with $r$, which
corresponds to the single-vortex region I of the ``phase diagram''.
We also note that the increase of the nonuniform evaporation rate
$J_s(r)=J_0(\theta)(1-r^2/R^2)^{-\lambda(\theta)}$ near the contact line
becomes more pronounced with decreasing the contact angle,
since $\lambda(\theta)=1/2-\theta/\pi$.
For this reason, at small contact angles the heat flow due to evaporative
cooling usually dominates
near the contact line and results in the reversed single-vortex
region III of the ``phase diagram''.
At large contact angles, the heat flow from the substrate
has much more chances to dominate, and this results in the
region I of the ``phase diagram''.

We also study the dependence of the fluid flows on the thickness
of the substrate. Fig.~\ref{diag} shows
the ``phase diagram'' obtained for different values of $h_R=h_S/R$.
As seen in Fig.~\ref{diag}, with an increase of the substrate thickness,
the subregion II$_2$ becomes dominating in the region II.
Further increase of the substrate thickness results in
shifting the regions I and II to larger contact angles.
The blue curve in Fig.~\ref{diag} is situated below
the black curve at small values of $h_R$,
while at larger values of $h_R$ it is above the black curve.
Therefore, the transition between the regions III and II$_2$
corresponds to the change of sign 
of the tangential component of the temperature gradient at the liquid--vapor interface
either at the droplet apex (for small substrate thickness) or
near the contact line (for larger substrate thickness).
The transition between the regions II$_2$ and II$_3$
corresponds to the change of sign 
of the tangential component of the temperature gradient at the liquid--vapor interface
either near the contact line (for small substrate thickness)
or at the droplet apex (for larger substrate thickness).
Fig.~\ref{plotABF} shows
the dependence of $\theta_{crit}$ on $h_R$ 
for the transitions between the regions I, II$_2$, II$_3$ and III.
A behavior of critical angles as a function of $h_R$ shown in Fig.~\ref{plotABF}
is, naturally, different in the region $h_R\lesssim 0.1$, where the blue
curve is below the black curve in Fig.~\ref{diag}, and in the region $h_R\gtrsim 0.1$,
where it is above the black curve.

\section{Conclusions}

The fluid flow structure in an evaporating sessile droplet of capillary size
has been considered in disregarding the convective heat transfer, i.e.,
for relatively small and slowly evaporating droplets
(when $Pe=\overline{u}R/\kappa \ll 1$).
It is shown that the region II of the ``phase diagram'' 
introduced in~\cite{Zhang} consists of the two subregions II$_2$ and II$_3$, where
II$_2$ corresponds to the flows with two vortices
and II$_3$ corresponds to the flows with three vortices. 
The transition between the regions III and II$_2$
corresponds to the change of sign 
of the tangential component of the temperature gradient at the liquid--vapor interface
either at the droplet apex (for small substrate thickness) or
near the contact line (for larger substrate thickness).
The transition between the regions II$_2$ and II$_3$
corresponds to the change of sign 
of the tangential component of the temperature gradient at the liquid--vapor interface
either near the contact line (for small substrate thickness)
or at the droplet apex (for larger substrate thickness).
Further increase of the substrate thickness
makes subregion II$_2$ dominating in region II,
and results in a shift of the regions I and II to larger contact angles.

Temperature distribution, fluid flow structure in the droplet
and general form of a ``phase diagram'' can be qualitatively
understood as a result of matching the
heat transfer through the solid--liquid interface
and the heat flow due to evaporative cooling.

The results may provide a better understanding of the Marangoni
effect of drying droplets and allow one to influence 
the fluid flow structure in evaporating droplets.
The results may potentially be useful to better 
understand numerous experimental results on
evaporative deposition patterns and self-assembly
of colloids and other materials.

\section*{Acknowledgements}

The simulations were partially carried out using facilities of
the Supercomputing Center of Lomonosov Moscow State University~\cite{Lomonosov}.
The results in Section~\ref{MethodsSec} were supported
by the Russian Science Foundation project No.~14-21-00158.

\end{document}